\documentclass[graybox]{svmult}

\usepackage{type1cm}        
\usepackage{makeidx}         
\usepackage{graphicx}        
\usepackage{multicol}        
\usepackage[bottom]{footmisc}

\usepackage{newtxtext}       %
\usepackage[varvw]{newtxmath}       

\usepackage{bm}

\usepackage[square,comma,numbers,sort&compress]{natbib}
\usepackage[colorlinks=true,linkcolor=blue,citecolor=red]{hyperref}%
\usepackage{color}

\def\erf{\mathrm{erf}}
\def\erfc{\mathrm{erfc}}
\def\erfcx{\mathrm{erfcx}}
\def\X{\bm{X}}
\def\x{\bm{x}}
\def\k{\bm{k}}
\def\n{\bm{n}}
\def\R{\mathbb{R}}
\def\pa{\partial\Omega}

\def\C{\mathbb{C}}
\def\I{\mathbb{I}}
\def\P{\mathbb{P}}
\def\E{\mathbb{E}}

\def\M{\mathcal{M}}
\def\L{\mathcal{L}}
\def\T{\mathcal{T}}

\makeindex             

\begin{document}

\title*{Encounter-based approach to target search problems: a review}
\author{Denis S. Grebenkov}
\institute{Denis S. Grebenkov \at 
Laboratoire de Physique de la Mati\`{e}re Condens\'{e}e,
CNRS -- Ecole Polytechnique, Institut Polytechnique de Paris, 91120 Palaiseau, France, 
\email{denis.grebenkov@polytechnique.edu}}
%
%
\maketitle

\abstract{
In this review, we present the encounter-based approach to target
search problems, in which the diffusive dynamics is described by the
joint probability of the position of the particle and the number of
its encounters with a given target set.  The knowledge of the
statistics of encounters allows one to implement various mechanisms of
reactions on the target set, beyond conventional reaction schemes.  We
formulate this approach for three relevant settings: discrete random
walks, Brownian motion with bulk reactions, and reflected Brownian
motion with surface reactions.  In all cases, we discuss the
advantages of this approach, its recent applications and possible
extensions. }

\section{Introduction}
\label{sec:1}

Marian von Smoluchowski formulated and solved one of the first target
search problems in chemical physics \cite{Smoluchowski17}.  He studied
the dynamics of colloidal particles diffusing in a solvent and
coagulating upon their encounters.  At the level of single-particle
events, his solution can be interpreted in terms of the survival
probability that two particles have not coagulated up to time $t$,
i.e., the random first-encounter time $\T$ does not exceed $t$.
Assuming that both particles are spheres of radii $R_1$ and $R_2$ that
undergo independent Brownian motions with diffusion coefficients $D_1$
and $D_2$, Smoluchowski associated the origin of the coordinate system
with the center of one particle and thus mapped the original
first-encounter problem for two moving particles onto an equivalent
target problem when a single point-like particle with diffusion
coefficient $D = D_1+D_2$ searches for a target -- the static sphere
of radius $R = R_1+R_2$.  As ordinary diffusion is governed by the
Laplace operator $\Delta$, Smoluchowski solved the diffusion equation
$\partial_t S_\infty(t|\x_0) = D \Delta S_\infty(t|\x_0)$ and found
the survival probability:
\begin{equation}  \label{eq:Sinf_ball}
S_\infty(t|\x_0) = \P_{\x_0} \{ \T > t\} = 1 - \frac{R}{|\x_0|} \erfc\biggl(\frac{|\x_0|-R}{\sqrt{4Dt}}\biggr),
\end{equation}
where $\erfc(z)$ is the complementary error function, $\x_0$ is the
initial position of the searcher in the target problem, while $|\x_0|$
can be interpreted as the initial distance between the centers of two
particles in the original problem.  Expectedly, the survival
probability is higher when the particles are initially farther from
each other, and decreases with time.  Curiously, the survival
probability does not vanish in the limit $t\to\infty$ because the
particle may escape to infinity without encountering the target due to
the transient character of Brownian motion in three dimensions.  

The seminal work by Smoluchowski emphasized the importance of
diffusion in chemical reactions that are nowadays called
diffusion-limited, diffusion-controlled, diffusion-influenced,
diffusion-assisted or diffusion-mediated reactions
\cite{Rice,Grebenkov23n}.  This field goes far beyond the original
coagulation dynamics and finds applications in physics, chemistry,
biology, ecology and social sciences
\cite{VanKampen,Redner,House,Schuss,Metzler,Lindenberg}.  The
particle can be an atom, an ion, a molecule, a protein, a virus, a
bacterium, or even an animal or a human.  The target can be another
particle, an enzyme, a receptor, a specific region on the DNA, on the
plasma membrane or inside the nucleus, a channel, a catalytic site, a
magnetic impurity, a trap, a hole, a pray, or a website.  The dynamics
can be an ordinary diffusion in the bulk, a continuous-time random
walk, a L\'evy flight, a biased spatially heterogeneous diffusion,
diffusion in a dynamic heterogeneous medium, or any other stochastic
process.  Finally, the reaction event can be a chemical
transformation, a binding to the adsorbing surface, an association to
another molecule, an escape or an entrance, a level crossing, a
relaxation of an excited state (e.g., loss of fluorescence or
transverse spin magnetization), eating, knowledge acquisition, or
death.  Since the influence of the dynamics onto the target problem
has been thoroughly investigated in the past, we keep the diffusive
dynamics as simple as possible and focus on different mechanisms of
reaction events.

In the original work by Smoluchowski, the coagulation of two particles
was assumed to occur immediately upon their first encounter.  This
condition was implemented via Dirichlet boundary condition to the
diffusion equation: $S_\infty(t|\x_0) = 0$ on the contact surface
(i.e., on the target).  In probabilistic terms, it simply states that
if the particle starts on the target, the target is immediately found
so that $\T = 0$ and the survival probability for any $t > 0$ is zero.
In many applications, however, the first encounter is not sufficient
for the reaction event \cite{Grebenkov23n}.  In fact, a diffusing
molecule and/or a target may need to be in appropriate conformational
states to bind or to overcome an energy activation barrier to react;
an ion channel needs to be in active (open) state to ensure the ion
transfer; the diffusing particle has to arrive onto an active
catalytic germ on the catalytic surface to be chemically transformed,
etc.  If such a condition is not fulfilled at the first arrival of the
particle onto the target, its diffusion is resumed until the next
arrival onto the target, and so.  As a consequence, the successful
reaction event is generally preceded by a sequence of failed reaction
attempts and the consequent diffusive explorations of the bulk around
the target.  This issue was recognized in 1949 by Collins and Kimball
how suggested to account for partial reactivity of the target via
Robin boundary condition (also known as radiation or third boundary
condition):
\begin{equation}  \label{eq:S_Robin}
- D \partial_n S_q(t|\x_0) = \kappa S_q(t|\x_0) ,
\end{equation}
where $\partial_n$ is the normal derivative oriented outwards the
confining domain, $\kappa$ is the reactivity of the boundary, and $q =
\kappa/D$ \cite{Collins49}.  This condition {\it postulates} that the
diffusive flux of particles from the bulk (the left-hand side) is
proportional to their reactive flux on the target (the right-hand
side) at each boundary point.  The reactivity $\kappa$ has units of
meter per second and ranges from $0$ for an inert impermeable target
(no reaction) to $+\infty$ for a perfectly reactive target, in which
case Dirichlet boundary condition is retrieved.  The emergence of
Robin boundary condition was further rationalized via various
microscopic mechanisms, including stochastic gating or rapid switching
between active and passive states in time
\cite{Benichou00,Reingruber09,Lawley15}, homogenization of
micro-heterogeneity of reactive patches
\cite{Berg77,Berezhkovskii04,Berezhkovskii06,Muratov08,Bernoff18,Grebenkov19b,Punia21},
overcoming energetic or entropic barriers or interactions
\cite{Weiss86,Hanggi90,Zhou91,Reguera06,Chapman16}.  Probabilistic
interpretations of Robin boundary condition via random walks and
reflected Brownian motions were discussed 
\cite{Grebenkov06,Grebenkov07,Erban07,Singer08,Grebenkov20,Piazza22,Grebenkov23b}.  
The role of reactivity onto diffusion-controlled reactions and related
target problems has been thoroughly investigated
\cite{Sano79,Brownstein79,Powles92,Sapoval94,Sapoval02,Grebenkov05,Traytak07,Bressloff08,Grebenkov15,Serov16,Grebenkov17a,Grebenkov18,Grebenkov18b,Grebenkov19c,Piazza19,Guerin21}.

The conventional approach for studying such diffusion-controlled
reactions relies on the analysis of the Laplace operator $\Delta$ that
governs diffusion in a given confining domain $\Omega$ with boundary
$\pa$.  For instance, when the confinement $\Omega$ is bounded, the
Laplace operator is known to have a discrete spectrum that allows one
to expand the survival probability and related quantities as
\cite{Redner,Gardiner}
\begin{equation}  \label{eq:Sq_spectral}
S_q(t|\x_0) = \sum\limits_{k=0}^\infty e^{-Dt\lambda_k^{(q)}}  u_k^{(q)}(\x_0) \int\limits_{\Omega} d\x \, u_k^{(q)}(\x),
\end{equation}
where $\lambda_k^{(q)}$ and $u_k^{(q)}(\x)$ are the eigenvalues and
normalized eigenfunctions satisfying:
\begin{equation}
-\Delta u_k^{(q)} = \lambda_k^{(q)} u_k^{(q)}  \quad \textrm{in}~\Omega, 
\qquad \partial_n u_k^{(q)} + q u_k^{(q)} = 0 \quad \textrm{on}~ \pa.
\end{equation} 
The superscript $q = \kappa/D$ highlights the dependence of
eigenvalues and eigenfunctions on the reactivity through Robin
boundary condition.
In this way, one can investigate the short-time and long-time behavior
of the survival probability in general domains, and find explicit
solutions in simple domains such as an interval, a rectangle, a disk,
a sphere, a circular annulus or a spherical shell
\cite{Redner,Gardiner,Grebenkov13}.  Despite the long history of
successful applications of this spectral method, it has several
drawbacks that hindered further advances in this field: (i) both
$\lambda_k^{(q)}$ and $u_k^{(q)}$ exhibit {\it implicit} dependence on
$q$ that hides the impact of the reactivity $\kappa$ on
diffusion-controlled reactions; for instance, if these quantities have
to be found numerically, the computation should be repeated for each
value of $q$; (ii) as the reactivity enters through Robin boundary
condition, the diffusive dynamics is tightly coupled to imposed
surface reactions in $S_q(t|\x_0)$; in other words, it is hard to
disentangle these effects; (iii) at the microscopic level, Robin
boundary condition describes surface reaction, which may occur at each
encounter with the target with a constant probability (see below);
this description does not allow to deal with more sophisticated
reaction mechanisms such as progressive passivation or activation of
the target by its interactions with the diffusing particle; (iv) when
diffusion occurs in an unbounded domain (as, e.g., in the original
Smoluchowski problem with $\Omega$ being the exterior of a ball of
radius $R$), the spectrum of the Laplace operator is continuous, and
the spectral expansion (\ref{eq:Sq_spectral}) is not applicable.

In this review, we describe an alternative approach to target search
problems that resolves all these limitations and provides
complementary insights onto diffusion-controlled reactions.  This
so-called encounter-based approach was first introduced in
\cite{Grebenkov20} and then further elaborated and extended in \cite{Grebenkov19a,Grebenkov20b,Grebenkov20c,Grebenkov21,Grebenkov22a,Bressloff22d,Benkhadaj22,Grebenkov22b,Bressloff22b,Grebenkov22c,Bressloff22a,Bressloff22c,Grebenkov23a,Bressloff23a,Bressloff23b}. 
In section \ref{sec:encounter}, we formulate this approach for three
settings: discrete random walks, Brownian motion with bulk reactions,
and reflected Brownian motion with surface reactions.  The first
setting is intuitively more appealing, even though exact computations
turn out to be the most difficult.  In turn, the formulation in the
last setting that relies on the notion of the boundary local time, is
more subtle but computations are usually easier.  Section
\ref{sec:spectral} briefly describes these computations via
spectral expansions based on the Dirichlet-to-Neumann operator.  In
section \ref{sec:applications}, we discuss several applications and
extensions of the encounter-based approach, whereas section
\ref{sec:conclusion} concludes the review.

\section{Probabilistic insights}
\label{sec:encounter}

In this section, we formulate the encounter-based approach for three
related settings and give some examples to illustrate the introduced
concepts.

\subsection{Discrete random walks}
\label{sec:discrete}

\begin{figure}[t!]
\begin{center}
\includegraphics[width=38mm]{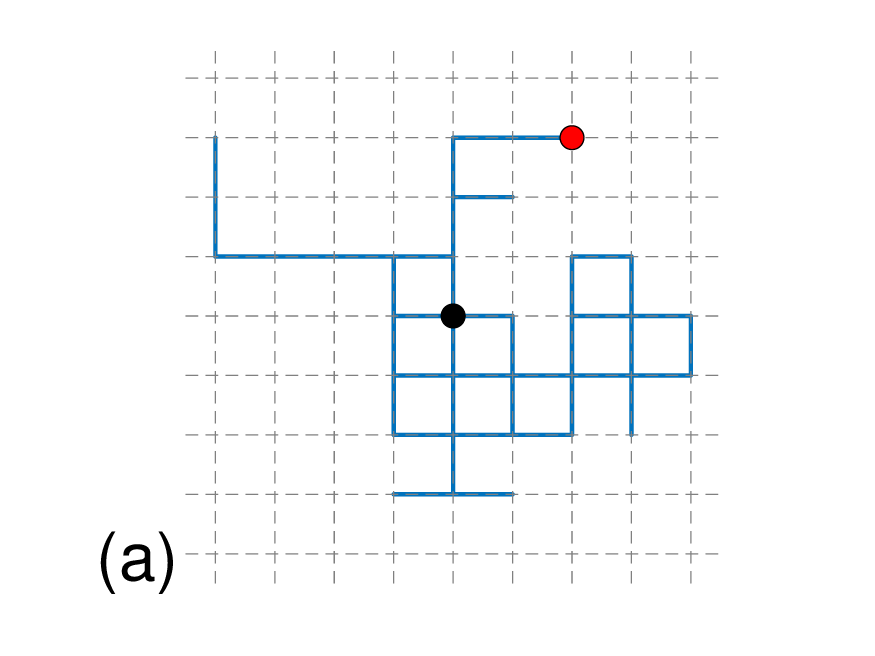} 
\includegraphics[width=38mm]{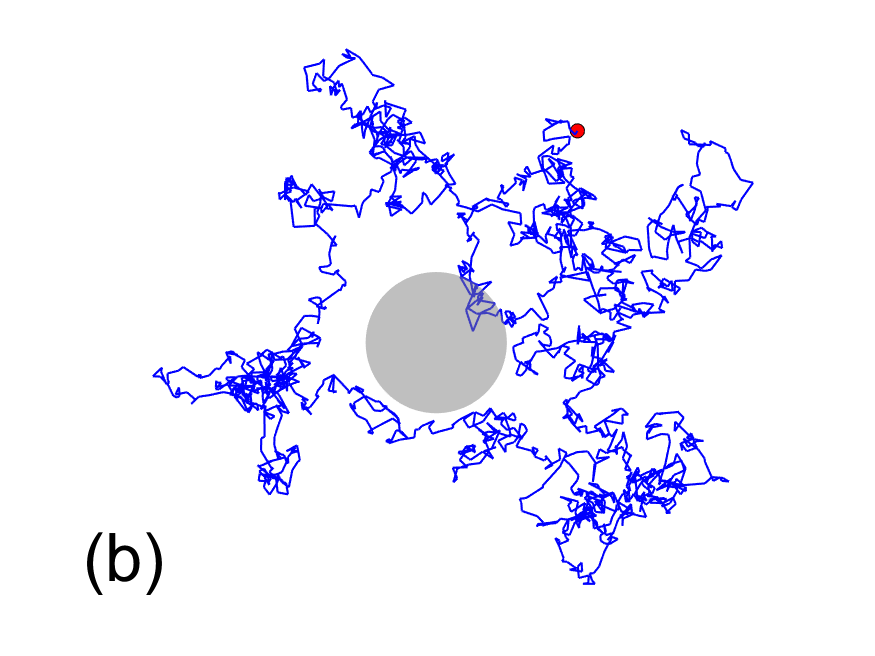} 
\includegraphics[width=38mm]{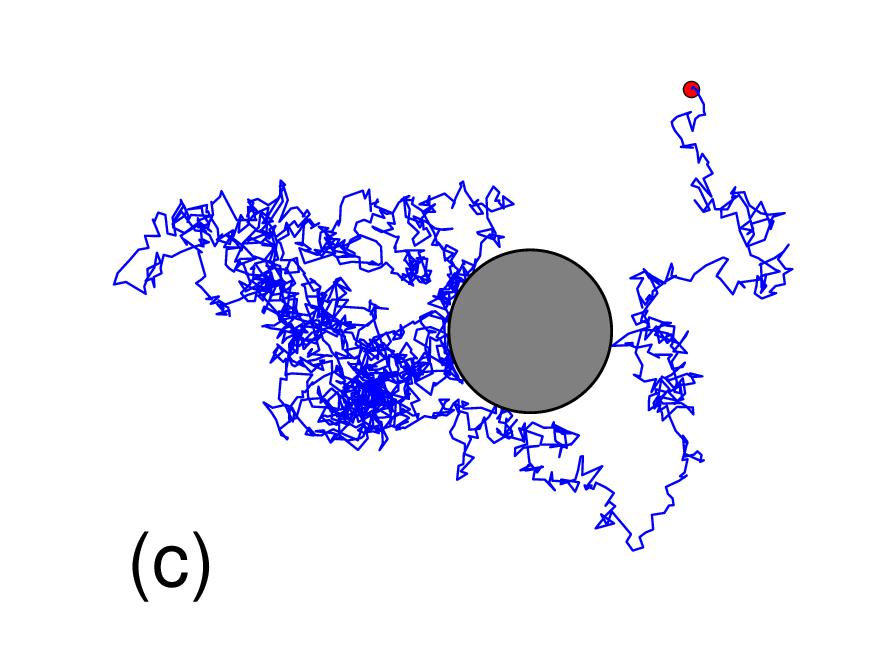} 
\end{center}
\caption{
A random trajectory (in blue) of a particle started from $\x_0$ (small
red disk) and searching for a target set. {\bf (a)} A random walk on
the square lattice visiting a single target site at the origin (small
black disk): $\Gamma = \{0\}$; {\bf (b)} Brownian motion that may
react when diffusing through the target set $\Gamma = \{ |\x| < R\}$
(gray disk); here, the particle dynamics is not affected by the
target; {\bf (c)} reflected Brownian motion that may react when
hitting the target set $\Gamma = \{ |\x| = R\}$ (in black) on the
boundary of an impenetrable obstacle (gray disk); the latter modifies
the trajectory of the particle.}
\label{fig:traj2d}
\end{figure}

To convey the main idea of the encounter-based approach, we start with
a simpler setting of a symmetric discrete-time random walk on a square
(or hyper-cubic) lattice with the spacing $a$ between the closest
nodes (Fig. \ref{fig:traj2d}(a)).  A walker starts from a point $\x_0$
and performs consecutive random jumps from its current position to one
of four neighboring sites with probability $1/4$.  For a given set
$\Gamma$ of target sites, we count the number of encounters of the
walker with the target set $\Gamma$ after $n$ steps,
\begin{equation}  \label{eq:Ln_def}
L_n = \sum\limits_{i=1}^n \I_{\Gamma}(\x_i),
\end{equation}
where $\x_i$ is the random position of the walker after $i$ steps, and
$\I_{\Gamma}(\x)$ is the indicator function of $\Gamma$:
$\I_{\Gamma}(\x) = 1$ if $\x\in\Gamma$ and $0$ otherwise.  In other
words, at each intermediate step $i = 1,2,\ldots,n$, the counter $L_n$
is incremented by $1$ if the walker's position $\x_i$ belongs to the
target set $\Gamma$ (i.e., the walker visits any site of $\Gamma$).
The random variable $L_n$ plays the central role in the
encounter-based approach.  We describe the diffusive dynamics of this
walker by the so-called full propagator
\begin{equation}
p(\x,L,n|\x_0) = \P_{\x_0}\{ \x_n = \x, ~ L_n = L\},
\end{equation}
i.e., the joint probability distribution of the walker's position
$\x_n$ and of the number of encounters $L_n$.  Rubin and Weiss
proposed a formal method to determine this quantity for any finite
configuration of target sites \cite{Rubin82}.  To avoid technical
issues, we skip the description of their method and just provide an
example below.  In turn, we discuss the advantages of the full
propagator $p(\x,L,n|\x_0)$ in describing target search problems.

In this framework, the walker merely counts its encounters with the
target set during the diffusive exploration of the space, i.e., the
full propagator $p(\x,L,n|\x_0)$ does not include any ``trigger'' to
stop this search process and to say that the target is found.  For
instance, if the target set can trap or bind the walker after some
encounters, such a stopping condition has to be added explicitly.  The
most common assumption is that the walker reacts with the target set
with a given probability $\sigma$ at each visit, and each reaction
attempt is independent from the dynamics and other reaction attempts.
The probability of finding the {\it survived} walker in $\x$ after $n$
steps, i.e., {\it without} reaction on the target set, follows
immediately as
\begin{equation}  \label{eq:g_p}
g_{\sigma}(\x,n|\x_0) = \sum\limits_{L=0}^\infty (1-\sigma)^L\, p(\x,L,n|\x_0).
\end{equation}
In fact, this sum classifies the contributions from different random
trajectories from $\x_0$ to $\x$ according to the number of encounters
with the target set.  For instance, $p(\x,0,n|\x_0)$ is the
probability of moving from $\x_0$ to $\x$ in $n$ steps without any
encounter with the target set.  The next term, $(1-\sigma)
p(\x,1,n|\x_0)$, is the contribution of trajectories that experienced
only one encounter with $\Gamma$, and this reaction attempt is failed
with probability $1-\sigma$.  Similarly, $(1-\sigma)^L p(\x,L,n|\x_0)$
is the contribution of trajectories that experienced $L$ encounters
with $\Gamma$, and each of these reaction attempts is failed, yielding
the factor $(1-\sigma)^L$.  Note that the sum in Eq. (\ref{eq:g_p})
has actually a finite number of terms because $p(\x,L,n|\x_0) = 0$ for
any $L > n$.
While $p(\x,L,n|\x_0)$ characterized the search process alone (without
any reaction), the propagator $g_\sigma(\x,n|\x_0)$ incorporates
reactions on the target set and thus implements the survival of the
walker.  The expression (\ref{eq:g_p}) relates these two fundamental
quantities and allows one to translate one from the other.

Conventional approaches to target search problems with random walks
deal directly with the propagator $g_{\sigma}(\x,n|\x_0)$ (or related
quantities), which is often easier to access than $p(\x,L,n|\x_0)$.
In fact, the propagator $g_{\sigma}(\x,n|\x_0)$ satisfies the master
equation
\begin{equation}
g_{\sigma}(\x,n+1|\x_0) = \frac{1}{2d} \left(\sum\limits_{|\x'-\x|=a} g_{\sigma}(\x',n|\x_0)\right) 
- \sigma \,\I_\Gamma(\x) g_{\sigma}(\x,n|\x_0),
\end{equation}
with the initial condition $g_{\sigma}(\x,0|\x_0) = \delta_{\x,\x_0}$,
which can be solved by standard tools \cite{Feller,Spitzer,Hughes}.
One may thus wonder what is the purpose of introducing a less
accessible quantity $p(\x,L,n|\x_0)$?  {\it Its disentanglement from
the reaction mechanism} is the fundamental advantage of the full
propagator, allowing one to consider more sophisticated reactions.
To illustrate this idea, let us assume that reaction on the target set
occurs after a prescribed number of encounters $L_{\rm max}$ (for
instance, a cleaning procedure or a security update is triggered when
the number of authorized visits exceeds $L_{\rm max}$).  The
conventional propagator $g_{\sigma}(\x,n|\x_0)$ does not allow one to
describe the position of the walker that is survived upon such a
stopping condition.  In turn, it is enough to sum the full propagator
$p(\x,L,n|\x_0)$ over $L$ from $0$ to $L_{\rm max}-1$ to determine the
probability of finding the walker at $\x$ at time $n$ before the
random walk is stopped at $L_{\rm max}$-th encounter.  More generally,
one can replace a fixed threshold $L_{\rm max}$ for the number of
encounters $L_n$ by a random threshold $\hat{L}$ that is defined by a
given probability law $\Psi(L) = \P\{ \hat{L} > L\}$.  In this case,
the reaction on the target set occurs at random moment $n$ when the
number of encounters $L_n$ reaches the random threshold $\hat{L}$.  In
other words, we employ the stopping condition $L_n = \hat{L}$ to
define the reaction mechanism, whereas the condition $L_n < \hat{L}$
describes the survival of the walker in such reactions.  As the
threshold $\hat{L}$ itself is random, the probability of finding the
survived walker at $\x$ after $n$ steps is
\begin{equation}  \label{eq:gPsi_p}
g_{\Psi}(\x,n|\x_0) = \P_{\x_0}\{ \x_n = \x, ~ L_n < \hat{L}\} = \sum\limits_{L=0}^\infty \Psi(L) \, p(\x,L,n|\x_0).
\end{equation}
Setting $\Psi(L) = \I_{[0,L_{\rm max}-1]}(L)$ yields the fixed
threshold at $L_{\rm max}$, whereas $\Psi(L) = (1-\sigma)^L$ implies
the earlier discussed conventional setting with a constant reaction
probability $\sigma$, see Eq. (\ref{eq:g_p}).  One sees that the
introduction of a random threshold opens the possibility to describe
much more general reaction mechanisms.  For instance, the constant
reaction probability $\sigma$ can be replaced by a sequence of
reaction probabilities $\sigma_1, \sigma_2, \ldots$, in which case the
contribution of the $L$-th term with $p(\x,L,n|\x_0)$ is weighted not
by $(1-\sigma)^L$, as in Eq. (\ref{eq:g_p}), but by
$(1-\sigma_1)(1-\sigma_2)\ldots(1-\sigma_L)$.  In other words, such a
model of encounter-dependent reactivity corresponds to
\begin{equation}
\Psi(L) = \prod\limits_{l=1}^L (1-\sigma_l).
\end{equation}
The controlled dependence of the reaction probability $\sigma_l$ on
the number of encounters allows one to describe activation or
passivation of the target set by its interactions with the random
walker.
Alternatively, one can rewrite this relation as $\Psi(L)/\Psi(L-1) = 1
- \sigma_L$ to determine the sequence of reaction probabilities
$\sigma_l$ that represents a given probability law $\Psi(L)$.  These
generalized reaction mechanisms were introduced in \cite{Grebenkov20}
for the encounter-based approach in the continuous setting (see
Sec. \ref{sec:cont_surface}).  We stress that such mechanisms require
the knowledge of the full propagator and therefore are not accessible
in conventional approaches.

\subsubsection*{Example of a single target site}

When the target set $\Gamma$ contains a single site, located at the
origin, the probabilities $p(\x,L,n|\x_0)$ can be determined from the
generating function
\begin{equation}
\tilde{p}(\x,L,z|\x_0) = \sum\limits_{n=0}^\infty z^n \, p(\x,L,n|\x_0).
\end{equation}
For a symmetric random walk on the square lattice, one gets
\cite{Montroll65,Rubin82}
\begin{equation}  \label{eq:tildep}
\tilde{p}(\x,L,z|\x_0) = \begin{cases}
\displaystyle \frac{\tilde{q}(\x,z) \tilde{q}(\x_0,z)}{[\tilde{q}(0,z)]^2} \biggl(1 - \frac{1}{\tilde{q}(0,z)}\biggr)^{L-1}  \qquad (L \geq 1), \cr
\displaystyle \tilde{q}(\x-\x_0,z) - \frac{\tilde{q}(\x,z) \tilde{q}(\x_0,z)}{\tilde{q}(0,z)} \hskip 11mm (L = 0), \end{cases}
\end{equation} 
where
\begin{equation}  \label{eq:tildeq_2D}
\tilde{q}(\x,z) = \frac{1}{(2\pi)^2} \int\limits_{-\pi}^{\pi} dk_1 \int\limits_{-\pi}^{\pi} dk_2  \, e^{-i(\x\cdot \k)/a}
\biggl(1 - z \frac{\cos(k_1) + \cos(k_2)}{2}\biggr)^{-1}
\end{equation} 
is the lattice Green's function, i.e., the generating function of the
probabilities $q(\x,n)$ of finding the walker at $\x$ after $n$ steps
on the square lattice without any target.  Note that $\tilde{q}(0,z)$
can be expressed as $\tilde{q}(0,z) = \frac{2}{\pi} K(z)$, where
$K(z)$ is the complete elliptic integral of the first kind
\cite{Guttmann10}.  From the propagator $p(\x,L,n|\x_0)$, one can
deduce other quantities of interest; e.g., Eq. (\ref{eq:g_p}) yields
the generating function for the propagator $g_{\sigma}(\x,n|\x_0)$:
\begin{equation} \label{eq:tildeg_sigma}
\tilde{g}_\sigma(\x,z|\x_0) = \sum\limits_{n=0}^\infty z^n \, g_\sigma(\x,n|\x_0) = \tilde{q}(\x-\x_0,z)
-  \frac{\sigma \, \tilde{q}(\x,z) \tilde{q}(\x_0,z)}{1 - (1-\sigma) \tilde{q}(0,z)} \,.
\end{equation}
In turn, the sum over $\x$ allows one to determine the statistics of
encounters, $\rho(L,n|\x_0) = \sum\nolimits_{\x} p(\x,L,n|\x_0)$, for
which we get for $L \geq 1$
\begin{equation}  \label{eq:tilde_rho}
\tilde{\rho}(L,z|\x_0) = \sum\limits_{n=0}^\infty z^n \, \rho(L,n|\x_0) = \sum\limits_{\x} \tilde{p}(\x,L,z|\x_0) 
= \frac{\bigl(1 - 1/\tilde{q}(0,z)\bigr)^{L-1} \tilde{q}(\x_0,z)}{(1-z) [\tilde{q}(0,z)]^2}  ,
\end{equation}
where we used the normalization of $q(\x,n)$ to evaluate the sum of
$\tilde{q}(\x,z)$ over $\x$.  After finding this generating function,
one can access these probabilities:
\begin{equation}  \label{eq:rho_n}
\rho(L,n|\x_0) = \P_{\x_0}\{ L_n = L\} = \frac{1}{n!} \lim\limits_{z\to 0} \frac{d^n}{dz^n} \tilde{\rho}(L,z|\x_0) .
\end{equation}
The same solution holds for the symmetric random walk on
higher-dimensional lattices with an appropriate modification of the
lattice Green's function $\tilde{q}(\x,z)$ \cite{Rubin82}.  For
instance, Joyce obtained an explicit representation for
$\tilde{q}(0,z)$ for the simple cubic lattice \cite{Joyce98}:
\begin{equation}
\tilde{q}(0,z) = \frac{(1-9\xi^4)\bigl[\frac{2}{\pi} K(\eta)\bigr]^2}{(1-\xi)^3 (1+3\xi)} ,  \quad
\eta = \sqrt{\frac{16\xi^3}{(1-\xi)^3 (1+3\xi)}}, \quad \xi = \sqrt{\frac{1-\sqrt{1-z^2/9}}{1+\sqrt{1-z^2}}} \,.
\end{equation}

Despite the explicit character of Eq. (\ref{eq:tildep}), a practical
implementation of this solution to get the probabilities
$p(\x,L,n|\x_0)$ is rather tedious even numerically.  In fact, one
needs to compute the double integral in Eq. (\ref{eq:tildeq_2D}) and
then to evaluate $n$-th order derivative of $\frac{d^n}{dz^n}
\tilde{p}(\x,L,z|\x_0)$ at $z = 0$.  Even the inversion of the fully
explicit relation (\ref{eq:tilde_rho}) with $\x_0 = 0$ to get
$\rho(L,n|0)$ is not straightforward, especially for large $n$.  In
turn, one can employ this formalism to study the asymptotic behavior.%
\footnote{
For instance, one can easily compute the generating function for the
mean number of encounters
\begin{eqnarray*}  \nonumber
f(z|\x_0) &=& \sum\limits_{n=0}^\infty z^n\, \E_{\x_0}\{ L_n \} = \sum\limits_{L=1}^\infty L\, \tilde{\rho}(L,z|\x_0) \\
&=& \frac{1}{1-z} \, \frac{\tilde{q}(\x_0,z)}{[\tilde{q}(0,z)]^2} \sum\limits_{L=1}^\infty L \bigl(1 - 1/\tilde{q}(0,z)\bigr)^{L-1}
= \frac{\tilde{q}(\x_0,z)}{1-z} \,.
\end{eqnarray*}
From the behavior of $f(z|\x_0)$ near $z = 1$, one can determine the
large-$n$ asymptotic behavior of the mean with the help of Tauberian
theorems.
%
For the simple cubic lattice, one has in the leading order $f(z|\x_0)
\approx \tilde{q}(\x_0,1)/(1-z)$ as $z\to 1$, so that $\E_{\x_0}\{ L_n
\} \to \tilde{q}(\x_0,1)$ as $n\to \infty$ (e.g., one has
$\tilde{q}(0,1) \approx 1.5164$ at $\x_0 = 0$).  Accounting for the
next-order correction allows one to show that the approach to this
limit is rather slow, as $1/\sqrt{n}$.  Note that the situation is
different for the square lattice, for which $\tilde{q}(0,1) \propto
\ln(1-z)$ so that $\E_{\x_0}\{ L_n \}$ exhibits a logarithmic growth
with $n$, due to recurrent nature of the random walk on the square
lattice.}
Moreover, one easily gets the first values of $\rho(L,n|0)$; e.g., for
the square lattice,
\begin{equation*}
\begin{array}{c | c c c c}  
\rho(L,n|0) & n=0 ~& n=2~ & n=4~   & n=6     \\  \hline
L=1       & 1   & 3/4 & 43/64 & 161/256 \\
L=2       & 0   & 1/4 & 17/64 &  69/256 \\
L=3       & 0   &  0  &  4/64 &  22/256 \\
L=4       & 0   &  0  &    0  &   4/256 \\  \end{array}
\end{equation*}  
and for the simple cubic lattice:
\begin{equation*}
\begin{array}{c | c c c c}  
\rho(L,n|0) & n=0 ~& n=2~ & n=4~   & n=6     \\  \hline
L=1       & 1   & 5/6 & 57/72 & 2995/3888 \\
L=2       & 0   & 1/6 & 13/72 &  731/3888 \\
L=3       & 0   &  0  &  2/72 &  144/3888 \\
L=4       & 0   &  0  &    0  &   18/3888 \\  \end{array}
\end{equation*}  
Each column gives the distribution of the number of encounters $L_n$
for even $n$ (for odd $n$, the distribution is the same as for $n-1$
because the walker needs an even number of steps to return to the
target site and to change $L_n$).  For instance, after two steps, the
walker started from the target site can either return to it (with
probability $(2d)^{-1}$) and thus increase the number of encounters to
$2$, or stay away from it and keep the initial number of encounters $L
= 1$ (with probability $1 - (2d)^{-1}$).

The above example illustrates that, even in the simplest case of a
single target site on the square lattice, the computation of the
probabilities $p(\x,L,n|\x_0)$ and $\rho(L,n|\x_0)$ is feasible but
rather sophisticated, even numerically.  When the number of target
sites increases, the computation is still feasible in a matrix form
(see \cite{Rubin82}), but the dependence of the propagators
$p(\x,L,n|\x_0)$ and $g_\sigma(\x,n|\x_0)$ on the starting and arrival
points and on the arrangement of the target sites becomes hidden behind
matrix operations.  Moreover, even though infinite lattices can be
replaced by general graphs to describe, e.g., random walks in confined
environments, the obtained results would be rather formal.

In the next section, we discuss how to generalize the encounter-based
approach to continuous diffusion, for which computations become
simpler.

\subsection{Brownian motion with bulk reactions}
\label{sec:cont_bulk}

In the continuous limit $a\to 0$, the discrete-step random walk $\x_n$
is known to approach Brownian motion $\X_t$, which is continuous in
time and space.  To extend the conventional approach to the continuous
setting, we consider now that the target set $\Gamma \subset\R^d$ as a
region in the bulk, and the particle diffusing inside $\Gamma$ can
undertake a first-order bulk reaction with the rate $Q$
(Fig. \ref{fig:traj2d}(b)).  The propagator $G_Q(\x,t|\x_0)$ satisfies
then the standard diffusion-reaction equation:
\begin{equation}  \label{eq:g_diff_I}
\partial_t G_Q(\x,t|\x_0) = D \Delta G_Q(\x,t|\x_0) - Q\, \I_{\Gamma}(\x) \, G_Q(\x,t|\x_0),
\end{equation}
subject to the initial condition $G_Q(\x,0|\x_0) = \delta(\x-\x_0)$
with the Dirac distribution $\delta(\x - \x_0)$ that fixes its
starting point $\x_0$.  Here the time evolution of $G_Q(\x,t|\x_0)$ is
governed by diffusion with a constant diffusion coefficient $D$ (first
term) and eventual disappearance due to bulk reactions inside $\Gamma$
(second term).
In contrast to the discrete setting, $G_Q(\x,t|\x_0)$ is not the
probability of arriving at $\x$ (which is strictly zero for continuous
diffusion), but the {\it probability density} of finding the particle
in a vicinity of $\x$ at time $t$, which is survived against bulk
reactions (see below).

It is worth noting that Szabo {\it et al.} proposed a more direct
``translation'' from the discrete to continuous setting by
incorporating point-like partial traps, localized at points
$\x_1,\ldots, \x_J$, via Dirac peaks into the diffusion equation
\cite{Szabo84}.  In other words, the reactivity profile
$\I_{\Gamma}(\x)$ in Eq. (\ref{eq:g_diff_I}) was replaced by
$\sum\nolimits_j \delta(\x-\x_j)$.  In this case, the propagator
$G_Q(\x,t|\x_0)$ in the presence of partial traps can be expressed in
terms of the propagator $G_0(\x,t|\x_0)$ without traps in a way, which
is similar to the discrete solution (\ref{eq:tildeg_sigma}).  However,
this approach is only applicable in the one-dimensional case%
\footnote{
The problem of localized partial traps was formulated in
\cite{Szabo84} in a way that might wrongly suggest that its solution
is valid in dimensions higher than $1$ (e.g., bold notation $\bm{r}$
for the position was employed).  This is not true, as the authors
stated on page 234: ``...but that in higher dimensions one must have a
reactive surface in order to have a properly posed problem.''  }
because a point is not accessible to Brownian motion in higher
dimensions.  In fact, the probability of visiting a given point (or
even a countable set of points) is strictly zero.  As a consequence, a
target site cannot be treated as point-like anymore but rather as a
small ball (or any open set $\Gamma \subset \R^d$ as we did above), to
make it accessible by Brownian motion.

To proceed towards an encounter-based approach, one can replace the
number of encounters $L_n$, introduced in Eq. (\ref{eq:Ln_def}) for a
random walk, by the residence time of Brownian motion inside $\Gamma$
up to time $t$:
\begin{equation}
L_t = \int\limits_0^t dt' \, \I_{\Gamma}(\X_{t'}) .
\end{equation}
One sees that the residence time is a specific functional of Brownian
trajectory $\X_t$.  According to the Feynman-Kac formula, the
propagator $G_Q(\x,t|\x_0)$ satisfying Eq. (\ref{eq:g_diff_I}) admits
the following probabilistic interpretation:
\begin{equation}  \label{eq:g_sigma_p_bulk}
G_Q(\x,t|\x_0) = \E_{\x_0} \biggl\{ e^{-Q L_t} \delta(\X_t - \x)\biggr\} 
= \int\limits_0^\infty dL \, e^{-QL} \, P(\x,L,t|\x_0).
\end{equation}
The first equality states that the probability density
$G_Q(\x,t|\x_0)$ accounts for all trajectories from $\x_0$ to $\x$ (of
duration $t$), whose contribution is weighted by their residence time
$L_t$ spent in $\Gamma$, during which the particle might react and
thus disappear with probability $1-e^{-QL_t}$.  In turn, the second
equality spells out the definition of this average in terms of the
full propagator $P(\x,L,t|\x_0)$, i.e., the joint probability density
of finding the particle, started from $\x_0$ at time $0$, in a
vicinity of $\x$ at time $t$, with the residence time in a vicinity of
$L$:
\begin{equation}
\P_{\x_0}\{ \X_t\in (\x,\x+d\x), L_t \in (L,L+dL)\} = P(\x,L,t|\x_0) d\x dL.
\end{equation}
If one manages to solve Eq. (\ref{eq:g_diff_I}) for a given target set
$\Gamma$, the full propagator $P(\x,L,t|\x_0)$ can be formally
obtained via the inverse Laplace transform with respect to $Q$.
Alternatively, as multiplication by $Q$ in Eq. (\ref{eq:g_diff_I})
becomes the derivative by $L$ in the dual space with respect to the
Laplace transform, one can write directly the partial differential
equation for the full propagator:
\begin{equation}  \label{eq:dP}
\partial_t P(\x,L,t|\x_0) = D\Delta P(\x,L,t|\x_0) - \I_\Gamma(\x) \partial_L P(\x,L,t|\x_0),
\end{equation}
subject to the initial condition $P(\x,L,0|\x_0) = \delta(\x-\x_0)
\delta(L)$.  The second term can be interpreted as the probability
flux due to the increase of the residence time when the particle
diffuses inside the target set $\Gamma$.  Multiplying this equation by
$e^{-QL}$ and evaluating its integral over $L$ from $0$ to $+\infty$
via integration by parts, one retrieves Eq. (\ref{eq:g_diff_I}) for
the propagator $G_Q(\x,t|\x_0)$, given that $P(\x,+\infty,t|\x_0) =0$,
and $P(\x,0,t|\x_0) = 0$ for any $\x\in\Gamma$.

Let us look again at Eq. (\ref{eq:g_sigma_p_bulk}), in which the
factor $e^{-QL}$ can be interpreted as the survival condition $\{L_t <
\hat{L}\}$, where $\hat{L}$ is the exponentially distributed random
threshold for the residence time $L_t$ that determines the bulk
reaction.  The exponential law for this threshold is the consequence
of choosing the first-order bulk reaction inside the target set
$\Gamma$.  In analogy to Eq. (\ref{eq:gPsi_p}), one can introduce an
arbitrary probability law $\Psi(L) = \P\{ \hat{L} > L\}$ for the
random threshold $\hat{L}$, in which case the propagator
$G_Q(\x,t|\x_0)$ is substituted by the generalized propagator
\begin{align}  \nonumber
G_\Psi(\x,t|\x_0) & = \E_{\x_0}\{ \delta(\X_t - \x) \cdot \I_{L_t < \hat{L}}\} 
= \E_{\x_0}\{ \delta(\X_t - \x) ~ \Psi(L_t) \} \\   \label{eq:GPsi_P_bulk}
& = \int\limits_0^\infty dL \, \Psi(L) \, P(\x,L,t|\x_0),
\end{align}
where $\I_{L_t < \hat{L}}$ is the indicator function of the
probabilistic event $L_t < \hat{L}$.  When $\Psi(L) \ne e^{-QL}$, the
bulk reaction is not of the first-order, and the generalized
propagator does not satisfy Eq. (\ref{eq:g_diff_I}).  One deals
therefore with non-Markovian bulk reactions introduced via
non-exponential random threshold $\hat{L}$ and its distribution
$\Psi(L)$.

\subsubsection*{Example of a ball as the target set}

To illustrate this approach, let us consider a ball of radius $R$ as
the target set: $\Gamma = \{ \x \in \R^3 ~:~ |\x|<R\}$.  While the
propagator $G_Q(\x,t|\x_0)$ can be found in this case, we focus
on the survival probability,
\begin{equation}
S_Q(t|\x_0) = \int\limits_{\R^3} d\x \, G_Q(\x,t|\x_0),
\end{equation}
i.e., the probability that the particle has not reacted up to time
$t$.  Note that the integral of Eq. (\ref{eq:g_sigma_p_bulk}) over
$\x$ implies that
\begin{equation}
S_Q(t|\x_0) = \int\limits_0^\infty dL \, e^{-Q L} \, \rho(L,t|\x_0),
\end{equation}
i.e., the inverse Laplace transform of $S_Q(t|\x_0)$ with respect to
$Q$ determines the statistics of the residence time $L_t$.
As the solution is fairly standard, we only sketch the main steps.
The survival probability satisfies backward diffusion-reaction
equation, alike Eq. (\ref{eq:g_diff_I}), but with respect to the
starting point $\x_0$.  In turn, its Laplace transform with respect to
time $t$,
\begin{equation}
\tilde{S}_Q(p|\x_0) = \int\limits_0^\infty dt \, e^{-pt} \, S_Q(t|\x_0),
\end{equation}
satisfies
\begin{equation}  \label{eq:tildeSQ}
\bigl(p - D \Delta_{\x_0} + Q\, \I_{\Gamma}(\x_0)\bigr) \tilde{S}_Q(p|\x_0) = 1.
\end{equation}
The rotational symmetry of $\Gamma$ implies that $\tilde{S}_Q(p|\x_0)$
depends only on the radial coordinate $r = |\x_0|$ that considerably
simplifies its computations.  Substituting $\Delta = \partial_r^2 +
\frac{2}{r} \partial_r$ into Eq. (\ref{eq:tildeSQ}) and writing its
solutions separately for $r < R$ and for $r > R$, one then matches two
solutions to ensure the continuity of $\tilde{S}_Q(p|\x_0)$ and of its
radial derivative at $r = R$.  After simplifications, one gets
\begin{equation}  \label{eq:tildeSQ_ball}
\tilde{S}_Q(p|\x_0) = \begin{cases}
\displaystyle \frac{1}{p'} - (C-1)\frac{p'-p}{pp'} \, \frac{R \, \sinh(|\x_0|\sqrt{p'/D})}{|\x_0| \, \sinh(R\sqrt{p'/D})}  \quad (|\x_0| < R), \cr
\displaystyle \frac{1}{p} - C \frac{p'-p}{pp'} \frac{R}{|\x_0|} e^{-(|\x_0|-R)\sqrt{p/D}} \hskip 14mm (|\x_0| \geq R), \end{cases} 
\end{equation} 
where 
\begin{equation}
p' = p + Q , \qquad C = \frac{1 - \tanh(R\sqrt{p'/D})/(R\sqrt{p'/D})}{1 + \sqrt{p/p'} \tanh(R\sqrt{p'/D})} \,.
\end{equation}
As a consequence, the inverse Laplace transform of
$\tilde{S}_Q(p|\x_0)$ with respect to $p$ yields $S_Q(t|\x_0)$,
whereas another inverse Laplace transform with respect to $Q$ gives
access to $\rho(L,t|\x_0)$:
\begin{equation}  \label{eq:rho_bulk}
\rho(L,t|\x_0) = \L^{-1}_{Q,L} \bigl\{  \L^{-1}_{p,t} \bigl\{ \tilde{S}_Q(p|\x_0) \bigr\} \bigr\} .
\end{equation}
Despite the explicit form of $\tilde{S}_Q(p|\x_0)$ in
Eq. (\ref{eq:tildeSQ_ball}), the above representation of
$\rho(L,t|\x_0)$ is rather formal because it involves two inverse
Laplace transforms.  Since $\tilde{S}_Q(p|\x_0)$ in
Eq. (\ref{eq:tildeSQ_ball}) depends on $p$ and $p' = p + Q$, it is
convenient to shift $Q$ by $p$ and write
\begin{equation}
\rho(L,t+L|\x_0) = \L_{L,Q}^{-1} \bigl\{ L_{p,t}^{-1} \bigl\{  \tilde{S}_{Q-p}(p|\x_0)  \bigr\} \bigr\}.
\end{equation}
This shifts allows one to evaluate the 
inverse Laplace transform with respect to $p$ explicitly.  For
instance, for $\delta = |\x_0| - R > 0$, one gets
%
%
\begin{eqnarray} \nonumber
\rho(L,t+L|\x_0) & =& \biggl[1 - \frac{R}{|\x_0|} \erfc\biggl(\frac{\delta}{\sqrt{4Dt}}\biggr)\biggr] \delta(L)
+ \frac{\sqrt{D}}{|\x_0|} \erfc\biggl(\frac{\delta}{\sqrt{4Dt}}\biggr) \L_{L,Q}^{-1} \biggl\{ \frac{1}{A(Q)} \biggr\} \\  \nonumber
&+& \frac{R}{|\x_0|} e^{-\delta^2/(4Dt)} \L_{L,Q}^{-1} \biggl\{ \biggl(1 -\frac{\sqrt{D}}{R A(Q)} \biggr)  \biggl[
\frac{A(Q)}{Q \sqrt{\pi t}} \\
&+& \biggl(1- \frac{A^2(Q)}{Q}\biggr) \erfcx\biggl(\frac{\delta}{\sqrt{4Dt}} + \sqrt{t} A(Q) \biggr)\biggr]\biggr\} ,
\end{eqnarray}
where $A(Q) = \sqrt{Q}/\tanh(R\sqrt{Q/D})$, and $\erfcx(z) = e^{z^2}
\erfc(z)$ is the scaled complementary error function.  Expectedly, the
first term accounts for trajectories that never arrive on the target
set $\Gamma$ so that the residence time remains $0$; the coefficient
in front of it is simply $S_\infty(t|\x_0)$, as if the (perfectly
reactive) target set was on the boundary.  The other terms can be
evaluated numerically or used to to investigate the asymptotic
behavior or the moments of the residence time.%
\footnote{
In practice, it is more convenient to evaluate the derivatives of
$\tilde{S}_Q(p|\x_0)$ with respect to $Q$ at $Q = 0$.  For instance,
to compute the mean residence time, one finds
\begin{equation*}
- \left. \frac{\partial \tilde{S}_Q(p|\x_0)}{\partial Q} \right|_{Q = 0} 
= \frac{\sqrt{D}}{|\x_0| p^{5/2}} \biggl(\frac{1 + R\sqrt{p/D}}{1 + \tanh(R\sqrt{p/D})} - 1\biggr),
\end{equation*}
so that
\begin{equation*}
\E_{\x_0} \{ L_t \} = \frac{\sqrt{D}}{|\x_0|} \biggl( - \frac{t^{3/2}}{\Gamma(5/2)} 
+ \L^{-1}_{p,t}\biggl\{ \frac{R\sqrt{p/D}+1}{p^{5/2}(1 + \tanh(R\sqrt{p/D}))}\biggr\} \biggr).
\end{equation*}  
As short times, one has $\E_{\x_0} \{ L_t \} \approx \frac{R
t}{2|\x_0|} \bigl(1 - \frac{4\sqrt{D t}}{3\sqrt{\pi} R} \bigr)$, with
exponentially small corrections.  In turn, at long times, one gets
$\E_{\x_0} \{ L_t \} \approx \frac{R^3}{3D|\x_0|} \bigl(1 -
\frac{R}{\sqrt{\pi D t}} + O(1/t)\bigr)$,
i.e., the residence time reaches a constant because the particle can
escape to infinity and thus never return to the target set.}

In the above discussion, the particle was assumed to diffuse in the
whole space $\R^d$.  The description can be easily adapted to consider
diffusion in a confined medium or a reservoir $\Omega\subset\R^d$ by
imposing Neumann boundary condition on the reflecting wall $\pa$,
i.e., by requiring that the diffusive flux of particles across the
wall vanishes.  The inclusion of this boundary condition does not
change the above discussion but may render the computation of the full
propagator $P(\x,L,t|\x_0)$ more sophisticated.

\subsection{Reflected Brownian motion with surface reactions}
\label{sec:cont_surface}

In the above description of bulk reactions, the particle can freely
diffuse through the target set $\Gamma$, i.e., the particle dynamics
is not affected by $\Gamma$.  For some target problems, this
description naturally represents the physical process; for instance, a
laser beam can relax the excited state of fluorescent particles in a
given spot, without alterting their diffusion.  However, there are
many other applications when reaction occurs on (a subset of) the
boundary of an impenetrable obstacle or on the solid wall of the
confining domain.  For instance, this is the case of heterogeneous
catalysis when the particle has to arrive onto a catalytic germ
located on the surface of a solid catalyst.  Alternatively, the
particle can bind to or adsorb on adherent patches on the boundary but
cannot go through it.  Similarly, if there is an ion channel in the
otherwise reflecting wall, the ion may exit or escape the confining
domain through this target set on the boundary, or be reflected back.
In all these cases, the target set is located on the boundary, which
strongly affects the particle dynamics due to eventual reflections.
We speak therefore about {\it surface reactions} that will be the main
focus in the rest of this review.  More precisely, we consider
ordinary diffusion of a point-like particle inside a confining domain
$\Omega \subset \R^d$ with a boundary $\pa$, and assume that the
target set $\Gamma$ is located on the boundary, i.e., $\Gamma = \pa$
or $\Gamma \subset \pa$ (Fig. \ref{fig:traj2d}(c)).

The standard diffusion-reaction equation (\ref{eq:g_diff_I}) is not
applicable in this setting.  In fact, this equation holds only for the
interior (bulk) points $\x \in \Omega$, for which $\I_{\Gamma}(\x)
\equiv 0$.  From the probabilistic point of view, as the target set
$\Gamma$ is a $(d-1)$-dimensional surface, the fraction of time that
Brownian motion spent on it is strictly zero, i.e., $L_t \equiv 0$ for
any $t$.  A simple way to overcome this problem is to introduce a thin
layer near $\Gamma$ of thickness $a$, $\Gamma_a = \{ \x\in \Omega~:~
|\x - \Gamma| < a\}$, where $|\x-\Gamma|$ is the Euclidean distance
between $\x$ and the target set $\Gamma$.  For this target set, one
can still apply Eq. (\ref{eq:g_diff_I}) to get the propagator
$G_Q^{(a)}(\x,t|\x_0)$ (the superscript $(a)$ highlights its
dependence on $a$).  If we keep the reaction rate $Q$ fixed and set
$a\to 0$, the effect of the reactive layer $\Gamma_a$ will disappear,
as explained above.  In order to preserve its effect, one has to
enhance the reaction rate as the layer gets thinner.  Setting $Q = q
D/a$ with a constant $q \geq 0$ and taking the limit $a\to 0$, one can
achieve a nontrivial limit
\begin{equation}
G_q(\x,t|\x_0) = \lim\limits_{a\to 0} G_{qD/a}^{(a)}(\x,t|\x_0).  
\end{equation}
One can show (see \cite{Grebenkov20}) that $G_q(\x,t|\x_0)$ satisfies
the diffusion equation,
\begin{equation}
\partial_t G_q(\x,t|\x_0) = D \Delta G_q(\x,t|\x_0)  \quad (\x\in\Omega),
\end{equation}
subject to the initial condition $G_q(\x,0|\x_0) = \delta(\x-\x_0)$ to
fix the starting point $\x_0$, Robin boundary condition on the target
set $\Gamma$, and Neumann boundary condition on the remaining part of
the boundary:
\begin{equation}
- \partial_n G_q(\x,t|\x_0) = q\, G_q(\x,t|\x_0)  \quad (\x\in\Gamma),  \qquad \partial_n G_q(\x,t|\x_0) = 0 \quad (\x\in\pa\backslash \Gamma).
\end{equation}
Comparing with Eq. (\ref{eq:S_Robin}), one sees that the parameter $q$
determines the reactivity $\kappa = qD$ of the target set.  Most
importantly, the conventional propagator $G_q(\x,t|\x_0)$ depends on
$q$ {\it implicitly}, through Robin boundary condition.

In the same vein, one can rescale the residence time $L_t^{(a)}$
inside the layer $\Gamma_a$ by $a$ to ensure the existence of a
nontrivial limit of $L_t^{(a)}/a$.  In this way, the {\it boundary
local time} $\ell_t$ is introduced:
\begin{equation}
\ell_t = \lim\limits_{a\to 0} \frac{D}{a} L_t^{(a)} = \lim\limits_{a\to 0} \frac{D}{a} \int\limits_0^t dt' \, \I_{\Gamma_a}(\X_{t'}) 
\end{equation}
(here the multiplication by $D$ is done for convenience).  Despite its
name, $\ell_t$ has units of length (we also stress that $\ell_t$
should not be confused with a bulk point local time, see
\cite{Borodin,Majumdar05}).  In practice, one can think of $\ell_t$ as
the rescaled residence time in a thin layer near the target set, i.e.,
$L_t^{(a)} \approx a \ell_t/D$, for small enough $a$.  Since the time
needed to cross a thin layer of width $a$ is of the order of $a^2/D$,
the ratio $L^{(a)}/(a^2/D) \approx \ell_t/a$ can be interpreted as the
number of such crossings, which represents the number of encounters
between the particle and the target set $\Gamma$ for reflected
Brownian motion.

As the representation (\ref{eq:g_sigma_p_bulk}) is valid for the
target set $\Gamma_a$, one can rewrite it by changing the integration
variable $\ell = DL/a$ and taking the limit $a\to 0$ as
\begin{equation}  \label{eq:GP}
G_q(\x,t|\x_0) = \int\limits_0^\infty d\ell \, e^{-q\ell} \, P(\x,\ell,t|\x_0),
\end{equation}
where the full propagator $P(\x,\ell,t|\x_0)$ is introduced as the
rescaled limit of the former full propagator $P(\x, L, t|\x_0)$
discussed in Section \ref{sec:cont_bulk}:
\begin{equation}
P(\x,\ell,t|\x_0) = \lim\limits_{a\to 0}  \frac{a}{D} P_{\Gamma_a}(\x, \ell a/D, t|\x_0).
\end{equation}
As previously, the full propagator is the joint probability density of
finding the particle that started from $\x_0$ at time $0$, in a
vicinity of $\x$ at time $t$, with the boundary local time $\ell_t$:
\begin{equation}
\P_{\x_0}\{ \X_t \in (\x,\x+d\x), ~ \ell_t \in (\ell,\ell+d\ell)\} = P(\x,\ell,t|\x_0) d\x d\ell .
\end{equation}

As earlier, this is the main building block of the encounter-based
approach.  In particular, the full propagator determines the
conventional propagator $G_q(\x,t|\x_0)$ via the Laplace transform in
Eq. (\ref{eq:GP}) with respect to the boundary local time $\ell$, in
direct analogy with Eq.  (\ref{eq:g_sigma_p_bulk}).  The major
advantage of the full propagator is that it describes the diffusive
motion {\it alone}, whereas the effect of surface reactions is taken
into account via the {\it explicit} factor $e^{-q\ell}$ in
Eq. (\ref{eq:GP}).  The disentanglement of diffusive dynamics from
surface reactions opens efficient ways for solving optimization
problems and introducing new surface reaction mechanisms.  In fact, as
in Sections \ref{sec:discrete} and \ref{sec:cont_bulk}, one can assume
that surface reaction occurs at time
\begin{equation}  \label{eq:Tdef}
\T = \inf\{ t > 0 ~:~ \ell_t > \hat{\ell}\}
\end{equation}
when the number of encounters, here represented by $\ell_t$, exceeds a
random threshold $\hat{\ell}$ with a given probability law $\Psi(\ell)
= \P\{ \hat{\ell} > \ell\}$.  As a consequence, the generalized
propagator for the survived particle reads
\begin{equation}
G_{\Psi}(\x,t|\x_0) = \E_{\x_0}\{ \delta(\X_t - \x) \cdot \I_{t < \T}\}
= \int\limits_0^\infty d\ell \, \Psi(\ell) \, P(\x,\ell,t|\x_0),
\end{equation}
in analogy to Eqs. (\ref{eq:gPsi_p}, \ref{eq:GPsi_P_bulk}).
Comparison with Eq. (\ref{eq:GP}) indicates that an exponentially
distributed threshold with $\Psi(\ell) = e^{-q\ell}$ describes the
conventional setting when a particle attempts to react at each
encounter with the target set $\Gamma$ with a constant reactivity
$\kappa = qD$.  In turn, other choices of $\Psi(\ell)$ correspond to
encounter-dependent reactivity and allow one to describe activation or
passivation of the target set (see further discussions in
\cite{Grebenkov20}).  Note that the generalized propagator
$G_{\Psi}(\x,t|\x_0)$ does not satisfy Robin boundary condition.

We complete this section by clarifying the following point.  To
provide more intuitive views on the encounter-based approach, we
started this section from the discrete case, for which the problem
could be easily formulated but getting its explicit solution was very
difficult.  Its extension to the continuous case was straightforward
for bulk reactions but less intuitive for surface reactions.  In
particular, we had to rescale the residence time to characterize the
encounters with the boundary target set $\Gamma$ via the boundary
local time $\ell_t$.  Such an indirect introduction of the
encounter-based approach for surface reactions was meant to ease its
presentation.  At the same time, it is important to stress that the
full propagator $P(\x,\ell,t|\x_0)$ emerges naturally and
straightforwardly in the mathematical theory of stochastic
differential equations in confined domains.  In fact, the mathematical
definition of reflected Brownian motion relies on the Skorokhod
equation \cite{Ito,Freidlin},
\begin{equation}
d\X_t = \sqrt{2D}\, d\bm{W}_t - \n(\X_t) \, d\ell_t ,
\end{equation}
where $\bm{W}_t$ is the standard Wiener process, and $\n(\x)$ is the
unit normal vector at a boundary point $\x$ oriented outwards the
domain $\Omega$.  The first term determines random displacements
inside $\Omega$, whereas the second term, which is nonzero only when
$\X_t$ is on the boundary $\pa$, describes normal reflections on $\pa$
and ensures that the process $\X_t$ does not leave the domain.
Curiously, this single equation determines simultaneously two tightly
related stochastic processes, the position $\X_t$ and the boundary
local time $\ell_t$.  In this setting, the full propagator
$P(\x,\ell,t|\x_0)$ naturally describes the pair $\{\X_t,\ell_t\}$.
This construction and other advantages of the encounter-based approach
were discussed in \cite{Grebenkov20}.

\section{Spectral insights}
\label{sec:spectral}

While the full propagator $P(\x,\ell,t|\x_0)$ plays the central role
in the encounter-based approach, its computation is a difficult task.
For instance, the standard inversion of the Laplace transform in
Eq. (\ref{eq:GP}) employs the Bromwich integral,
\begin{equation}
P(\x,\ell,t|\x_0) = \frac{1}{2\pi i} \int\limits_{\gamma - i\infty}^{\gamma + i\infty} dq \, e^{q\ell} \, G_q(\x,t|\x_0),
\end{equation}
where $\gamma$ is a vertical contour in the complex plane $q\in\C$
chosen so that all singularities of $G_q(\x,t|\x_0)$ are to the left
of it.  In other words, it requires the knowledge of the conventional
propagator $G_q(\x,t|\x_0)$ even for complex values of the parameter
$q$ (which was supposed to be positive throughout this review).  In
practice, one needs other representations to access the full
propagator more efficiently.  The conventional approach employs
spectral expansions, alike Eq. (\ref{eq:Sq_spectral}), based on the
eigenfunctions and eigenvalues of the Laplace operator that governs
the diffusive dynamics \cite{Redner,Gardiner}.  In turn, the
encounter-based approach was shown to rely on another mathematical
operator, known as the Dirichlet-to-Neumann operator $\M_p$
\cite{Grebenkov20}.

To introduce this operator, let us consider an auxiliary steady-state
diffusion-reaction problem when particles diffuse inside a confining
domain $\Omega$ and can undergo first-order bulk reactions with a rate
$p$.  In the steady-state regime, their concentration $u(\x)$ obeys
the modified Helmholtz equation: $D\Delta u - p u = 0$.  If a
prescribed concentration of particles $f$ is maintained on the target
set $\Gamma \subset\pa$, one can think of a permanent source of
particles on $\Gamma$, which diffuse in $\Omega$ and disappear due to
bulk reactions.  The diffusive flux of particles started from
$\Gamma$, $g = D(\partial_n u)|_{\Gamma}$, is fully determined by the
solution $u$ of the modified Helmholtz equation, which, in turn, is
determined by the imposed concentration $f$.  In other words, to a
given concentration $f$ on $\Gamma$, one can associate the diffusive
flux density $g$ on $\Gamma$.  This is precisely the action of the
Dirichlet-to-Neumann operator $\M_p$.  More formally, for a given
function $f$ on $\Gamma$, one defines $\M_p f = (\partial_n
u)|_{\pa}$, where $u$ is the solution of the boundary value problem
\begin{equation}
(p - D\Delta) u = 0 \quad (\x\in\Omega), \qquad u|_{\Gamma} = f, \qquad (\partial_n u)|_{\pa\backslash \Gamma} = 0.
\end{equation} 
When $p\geq 0$ and the target set $\Gamma$ is bounded and smooth
enough, $\M_p$ is self-adjoint elliptic pseudodifferential operator
with a discrete spectrum (see technical details and mathematical
restrictions in \cite{Levitin} and references therein).  In other
words, there is a countable sequence of eigenvalues $\mu_k^{(p)}$ and
eigenfunctions $v_k^{(p)}$, enumerated by $k = 0,1,2,\ldots$, that
satisfy $\M_p v_k^{(p)} = \mu_k^{(p)} v_k^{(p)}$.  Each eigenfunction
$v_k^{(p)}$ living on $\Gamma$ can be extended into the whole domain
$\Omega$ as the unique solution $V_k^{(p)}$ of the boundary value
problem:
\begin{equation}
(p - D\Delta) V_k^{(p)} = 0 \quad (\x\in\Omega), \qquad V_k^{(p)}|_{\Gamma} = v_k^{(p)}, 
\qquad (\partial_n V_k^{(p)})|_{\pa\backslash \Gamma} = 0.
\end{equation} 
In mathematics, $V_k^{(p)}$ are called the eigenfunctions of the
Steklov spectral problem.  In analogy to the eigenfunctions
$u_k^{(q)}$ of the Laplace operator, one can use $V_k^{(p)}$ in
spectral expansions.  In particular, these eigenfunctions allowed us
to determine the full propagator
\cite{Grebenkov20}:
\begin{equation}  \label{eq:P_spectral}
P(\x,\ell,t|\x_0) = G_\infty(\x,t|\x_0) \delta(\ell) + \L^{-1}_{p,t}\biggl\{ \frac{1}{D} 
\sum\limits_{k=0}^\infty V_k^{(p)}(\x_0) V_k^{(p)}(\x) e^{-\mu_k^{(p)} \ell} \biggr\} ,
\end{equation} 
where $\L^{-1}_{p,t}$ denotes the inverse Laplace transform with
respect to time $t$.  The first term describes the contribution of
trajectories of duration $t$, that moved from $\x_0$ to $\x$ without
hitting the target set $\Gamma$.  The probability of this event is
determined by the conventional propagator $G_\infty(\x,t|\x_0)$ with
Dirichlet boundary condition $G_\infty(\x,t|\x_0)|_{\x\in\Gamma} = 0$
that prohibits any arrival onto $\Gamma$.  As these trajectories never
encountered $\Gamma$, the boundary local time remains $0$, as
expressed by the factor $\delta(\ell)$.  In turn, the second term
accounts for all other trajectories that could encounter the target
set $\Gamma$ and thus correspond to positive values $\ell$ of the
boundary local time $\ell_t$.

The spectral expansion (\ref{eq:P_spectral}) is the cornerstone of the
encounter-based approach.  Many other characteristics of search
processes and diffusion-controlled reactions can be expressed with the
help of this relation, including the conventional propagator
$G_q(\x,t|\x_0)$ and its extension $G_\Psi(\x,t|\x_0)$, the
surface-hopping propagator \cite{Grebenkov20b}, the distribution of
the boundary local time \cite{Grebenkov07,Grebenkov19a}, the survival
probability, the probability density of various first-passage times
\cite{Grebenkov20c}, the particle concentration, the spread harmonic
measure \cite{Grebenkov06b,Grebenkov15}, the overall reaction rate,
etc.  The intricate connection to the Dirichlet-to-Neumann operator
opens promising opportunities to translate the knowledge on this
operator from spectral geometry \cite{Levitin} to diffusion-controlled
reactions and related target search problems.

\subsubsection*{Example of a spherical target set}

The spectral expansion (\ref{eq:P_spectral}) is also an efficient
practical tool for computing the full propagator (see, e.g.,
\cite{Grebenkov20b,Grebenkov20c}).  For instance, if the target set
$\Gamma$ is the boundary of a ball of radius $R$, $\Gamma =
\{\x\in\R^3 ~:~ |\x| = R\}$, whereas the particle diffuses in the
exterior of this ball, $\Omega = \{\x\in\R^3 ~:~ |\x| > R\}$, the
eigenvalues $\mu_k^{(p)}$ and eigenfunctions $V_k^{(p)}$ have a
relatively simple form.  To skip technical details, let us focus again
on the statistics of encounters by integrating
Eq. (\ref{eq:P_spectral}) over $\x\in \Omega$.  In this case, the
first term in Eq. (\ref{eq:P_spectral}) is just the survival
probability of a particle outside a perfected reactive sphere, given
by Eq. (\ref{eq:Sinf_ball}), while the sum in the second term contains
only one term due to the rotational invariance of the problem.
Surprisingly, the inverse Laplace transform of this single term can be
computed explicitly, yielding%
\footnote{
Note that there is a misprint in Eq. (5) of \cite{Grebenkov21}:
$\erf(z)$ should be replaced by $\erfc(z)$, as in our
Eq. (\ref{eq:rhoell_ball}). }
\cite{Grebenkov21}
\begin{eqnarray}  \nonumber
\rho(\ell,t|\x_0) &=& \biggl(1 - \frac{R}{|\x_0|} \erfc\biggl(\frac{|\x_0|-R}{\sqrt{4Dt}}\biggr)\biggr) \delta(\ell)  \\  \label{eq:rhoell_ball}
&+& \frac{e^{-\ell/R}}{|\x_0|} \biggl(\erfc\biggl(\frac{|\x_0|-R+\ell}{\sqrt{4Dt}}\biggr) 
+ \frac{R e^{-(|\x_0|-R+\ell)^2/(4Dt)}}{\sqrt{\pi Dt}}\biggr).
\end{eqnarray}
To appreciate the remarkable simplicity of this result, it is enough
to compare it with similar quantities $\rho(L,n|\x_0)$ and
$\rho(L,t|\x_0)$ given by Eqs. (\ref{eq:rho_n}, \ref{eq:rho_bulk}).
Unfortunately, this is an exception: even for the circular target set
in the plane, the expression for $\rho(\ell,t|\x_0)$ is much more
involved \cite{Grebenkov21}.

We complete this discussion by stressing that the three
encounter-based approaches presented in Sections \ref{sec:discrete},
\ref{sec:cont_bulk}, and \ref{sec:cont_surface} are similar but not fully
equivalent.  Indeed, despite different units, the number of encounters
$L_n$, the residence time $L_t$ and the boundary local time $\ell_t$
convey essentially the same information about the interaction of the
diffusing particle with the target set.  For instance, the means of
these random variables, $\E_{\x_0} \{ L_n\}$, $\E_{\x_0}\{L_t\}$ and
$\E_{\x_0}\{ \ell_t\}$, reach constant values in the long time limit
for the search in three dimensions (or higher).  This is the
consequence of the transient character of the diffusive motion, which
can escape to infinity and never return to the target set.  Moreover,
this approach is quite slow,%
\footnote{
To get the mean boundary local time, it is easier to start from the
known expression for the survival probability
\begin{equation*}
\tilde{S}_q(p|\x_0) = \frac{1}{p} \biggl(1 - \frac{qR^2 e^{-(|\x_0|-R)\sqrt{p/D}}}{|\x_0|(1 + qR + R\sqrt{p/D})}\biggr).
\end{equation*}
Since this is the double Laplace transform of
Eq. (\ref{eq:rhoell_ball}), the mean boundary local time can be found
by evaluating the derivative of this expression with respect to $q$ at
$q = 0$ and then computing the inverse Laplace transform with respect
to $p$:
\begin{equation*}
\E_{\x_0}\{ \ell_t\} = \frac{R^2 e^{-(|\x_0|-R)^2/(4Dt)}}{|\x_0|} \biggl[\erfcx\biggl(\frac{|\x_0|-R}{\sqrt{4Dt}}\biggr)
- \erfcx\biggl(\frac{|\x_0|-R}{\sqrt{4Dt}} + \sqrt{Dt}/R\biggr)\biggr].
\end{equation*}
The mean boundary local time vanishes in the short-time limit and
approaches the constant $R^2/|\x_0|$ in the long-time limit because
the particle can escape to infinity and never return to the target set
with the probability $R/|\x_0|$.  More precisely, one has $\E_{\x_0}\{
\ell_t\} = R^2/|\x_0| - R^2/\sqrt{\pi Dt} + O(1/t)$ as $t\to\infty$.
If the particle starts on the target set, $|\x_0| = R$, one finds
$\E_{\x_0}\{\ell_t\} = R (1 - \erfcx(\sqrt{Dt}/R))$.  Similarly, one
can find the variance and higher-order moments of the boundary local
time.}
typically as $n^{-1/2}$ or $t^{-1/2}$.  At the same time, the
diffusive motion of the particle is hindered by the target set due to
reflections in the case of reflected Brownian motion, in sharp
contrast to bulk reactions.  As a consequence, the statistics of
random trajectories can be considerably different between the cases of
bulk and surface reactions.  Even if the target set is small, the
distributions of $L_t$ and $\ell_t$ differ.  In fact, as the residence
time $L_t$ is a fraction of time that Brownian motion spends on the
target set, it cannot exceed $t$, so that the distribution of $L_t$ is
supported on a finite interval $(0,t)$.  In turn, the boundary local
time $\ell_t$ can take any positive value on $\R_+$, even though the
probability of very high $\ell_t$ is extremely small, see
Eq. (\ref{eq:rhoell_ball}).

\section{Applications and extensions}
\label{sec:applications}

The encounter-based approach provided complementary insights onto
diffusion-controlled reactions and stimulated new developments in
first-passage phenomena in statistical physics.  In this section, we
list some of these achievements and highlight the related open
problems.

\subsubsection*{A variety of first-passage times}

In previous sections, we showed that the successful realization of a
reaction event occurs when the number of encounters exceeds some
threshold.  In particular, the random time of the reaction event can
be defined via Eq. (\ref{eq:Tdef}) as the first-crossing time of a
random threshold $\hat{\ell}$.  The distribution of reaction times,
which characterizes the efficiency of diffusion-controlled reactions,
was thoroughly investigated for conventional surface reactions with a
constant reactivity (see
\cite{Redner,Metzler,Lindenberg,Grebenkov18,Grebenkov18b,Grebenkov19c,Guerin21}
and references therein).  As the boundary local time $\ell_t$ is a
nondecreasing process, the survival condition $\T > t$ is identical to
$\ell_t < \hat{\ell}$, i.e., the particle is survived up to time $t$
if its boundary local time $\ell_t$ has not yet crossed the threshold
$\hat{\ell}$.  As a consequence, the survival probability, i.e., the
integral of the propagator, determines the distribution of the
reaction time $\T$:
\begin{eqnarray}  \nonumber
\P_{\x_0}\{ \T > t\} &=& \int\limits_{\Omega} d\x \, G_{\Psi}(\x,t|\x_0) 
= \int\limits_0^\infty d\ell \, \Psi(\ell) \int\limits_{\Omega} d\x \, P(\x,\ell,t|\x_0) \\
&=& \int\limits_0^\infty d\ell \, \Psi(\ell) \rho(\ell,t|\x_0).
\end{eqnarray}
In other words, the knowledge of the probability density
$\rho(\ell,t|\x_0)$ of the boundary local time $\ell_t$ determines the
distribution of reaction times.  For instance, setting $\Psi(\ell) =
e^{-q\ell}$, one can express the survival probability $S_q(t|\x_0)$,
which is a cornerstone of the conventional approach to first-passage
times, as the moment-generating function of the boundary local time:
\begin{equation}
S_q(t|\x_0) = \P_{\x_0}\{ \T_q > t\} = \int\limits_0^\infty d\ell \, e^{-q\ell} \, \rho(\ell,t|\x_0) = \E_{\x_0}\{ e^{-q\ell_t}\} .
\end{equation}
The latter can be obtained by integrating Eq. (\ref{eq:P_spectral})
over the confining domain $\Omega$:
\begin{equation}  \label{eq:rho_spectral}
\rho(\ell,t|\x_0) = S_\infty(t|\x_0) \delta(\ell) + \L^{-1}_{p,t}\biggl\{ \frac{1}{D} 
\sum\limits_{k=0}^\infty e^{-\mu_k^{(p)} \ell} V_k^{(p)}(\x_0)  \int\limits_{\Omega} d\x \, V_k^{(p)}(\x) \biggr\} .
\end{equation} 
While this analysis can be extended in several directions, we mention
only two of them.  

(i) One can consider $N$ independent particles searching
simultaneously for the target set $\Gamma$, and investigate various
extreme value statistics for this ensemble, such as, e.g., the minimum
of their reaction times $\T_q^1,\ldots,\T_q^N$.  Since the seminal
work by Weiss {\it et al.} \cite{Weiss83}, the distribution of the
{\it fastest} first-passage time and its mean were actively studied
for conventional surface reactions
\cite{Bray13,Schuss19,Basnayake19,Lawley20,Lawley20b,Lawley20c,Madrid20,Grebenkov20a,Grebenkov22}.
The encounter-based approach offers an alternative view onto this
problem \cite{Grebenkov22b}.  If $\ell_t^1, \ldots, \ell_t^N$ denote
the boundary local times of each of $N$ particles, one can define the
first crossing-time $\T_{\ell,N} = \inf\{ t > 0~:~ \ell_t^1 + \ldots +
\ell_t^N > \ell\}$ when the total number of encounters exceeds a given
threshold $\ell$.  If each encounter is associated with consumption of
resources on the target set $\Gamma$, the random variable
$\T_{\ell,N}$ determines the moment when the initial amount $\ell$ of
resources is depleted by a population of diffusing species.  The
distribution of $\T_{\ell,N}$ can be determined in terms of the
survival probability $S_q(t|\x_0)$ for a single particle: $\P_{\x_0}\{
\T_{\ell,N} > t\} = \L_{q,\ell}^{-1}\{ [S_q(t|\x_0)]^N/q\}$, from
which many asymptotic results can be deduced \cite{Grebenkov22b}.
Choosing the threshold $\ell = 0$ implies that none of the particles
has reached the target set, so that $\T_{0,N} = \min\{ \T_\infty^1,
\ldots, \T_\infty^N\}$, i.e., one retrieves the conventional fastest
first-passage time to the perfectly reactive target studied earlier.
Replacing $\ell$ by a random exponentially distributed threshold
$\hat{\ell}$, one gets $\T_{\hat{\ell},N} = \min\{ \T_q^1, \ldots,
\T_q^N\}$ for a partially reactive target (see \cite{Grebenkov22b} for
more details).  In turn, the choice of a random threshold $\hat{\ell}$
with an arbitrary distribution $\Psi(\ell)$ may yield more
sophisticated extreme value statistics.

(ii) Alternatively, one can consider many target sets, $\Gamma_1,
\ldots, \Gamma_n$, and investigate the statistics of their encounters
by a single particle.  For each target set $\Gamma_i$, one can
introduce its own boundary local time $\ell_t^i$ and then study the
full propagator $P(\x,\ell^1,\ldots,\ell^n,t|\x_0)$ as the joint
probability density of the position $\x$ and boundary local times
$\ell_t^1, \ldots, \ell_t^n$.  In this setting, different kinds of
first-passage times emerge, e.g., the first instance when the maximum
or the minimum of $\ell_t^1,\ldots,\ell_t^n$ exceeds some threshold,
or the first instance when one boundary local time exceeds the other,
and so on.  These questions naturally generalize the notion of
splitting probabilities that characterize diffusional screening or
diffusion interactions between the targets that compete for the
diffusing particle.  Some of these first-passage problems were
investigated in simple geometric settings with two target sets
\cite{Grebenkov20c}.  Moreover, if one of the target sets is perfectly
reactive, one deals with the escape problem \cite{Grebenkov23a}.  For
instance, one can investigate the statistics of encounters of a
protein with its receptor (a target set $\Gamma_1$) in the presence of
a protein channel (another target set $\Gamma_2$) on the plasma
membrane through which the protein can leave the living cell
\cite{Lauffenburger}.  Despite recent progress in this field, many
problems with multiple target sets are still open.  For instance, a
spectral expansion for the full propagator
$P(\x,\ell^1,\ldots,\ell^n,t|\x_0)$ is missing.

\subsubsection*{Small target limit}

According to the spectral expansion (\ref{eq:rho_spectral}), the
dependence of $\rho(\ell,t|\x_0)$ on the shape of the confining domain
$\Omega$ and the target set $\Gamma$ is captured implicitly via the
spectral properties of the Dirichlet-to-Neumann operator $\M_p$.  In
general, it is very hard to get the statistics of encounters
explicitly.  However, if the target set is small and located far away
from the remaining confining boundary $\pa\backslash \Gamma$, some
approximations can be derived.  This setting, known as the narrow
escape or narrow capture problem, was thoroughly investigated for
conventional surface reactions (see
\cite{Schuss07,Benichou08,Holcman14} and references therein).
Bressloff extended previous works by employing the matched asymptotic
analysis to obtain an approximate form of the full propagator
\cite{Bressloff22b}.  These results were compared to two other
approximations in \cite{Grebenkov22c}.  In particular, a fully
explicit approximation was derived in three dimensions for
$\rho(\ell,t|\x_0)$ averaged over the uniformly distributed starting
point $\x_0$:
\begin{eqnarray}  \nonumber
\overline{\rho(\ell,t)} &=& \frac{1}{|\Omega|} \int\limits_\Omega d\x_0 \, \rho(\ell,t|\x_0)  \\
& \approx & e^{-t/t_0} \delta(\ell) + \sqrt{\frac{t/t_0}{\ell \ell_0}} e^{-\ell/\ell_0 - t/t_0} 
I_1\biggl(2\sqrt{(t/t_0)(\ell/\ell_0)}\biggr),
\end{eqnarray}
where $|\Omega|$ is the volume of the confining domain $\Omega$,
$I_\nu(z)$ is the modified Bessel function of the first kind, $\ell_0
= |\Gamma|/C$, $t_0 = |\Omega|/(DC)$, $|\Gamma|$ is the surface area
of the target set, and $C$ is the harmonic capacity (or capacitance)
of the target set (e.g., $C = 4\pi R$ is the capacity of a ball of
radius $R$) \cite{Grebenkov22c}.  Moreover, when $t\ell \gg
t_0\ell_0$, the asymptotic behavior of the modified Bessel function
yields a remarkably simple expression:
\begin{equation}  
\overline{\rho(\ell,t)} \approx e^{-t/t_0} \delta(\ell) + 
\frac{(t/t_0)^{\frac14}}{2\sqrt{\pi} (\ell/\ell_0)^{\frac34} \ell_0} e^{-\bigl(\sqrt{\ell/\ell_0} - \sqrt{t/t_0}\bigr)^2} .
\end{equation}
As time increases, the particle experiences more encounters with the
target set, so that the maximum of this density is shifted to larger
$\ell$.  Despite the simple form of this expression, the dependence on
$\ell$ is quite nontrivial.  One also concludes that when the target
set is small, the statistics of encounters is essentially determined
by two scales, $t_0$ and $\ell_0$, that are formed by three geometric
characteristics: the surface area and the capacity of the target set,
and the volume of the domain.

\subsubsection*{Reversible binding}

The spectral expansion (\ref{eq:P_spectral}) is particularly well
suited for describing diffusive explorations of the confining domain
$\Omega$ between reflections on the target set $\Gamma$.  In
particular, the notion of the surface-hopping propagator, that
characterizes the position of the particle after a prescribed number
of encounters, was introduced in \cite{Grebenkov20b}.  Moreover, an
accurate description of repeated returns to the target set allows one
to consider reversible binding of the particle on $\Gamma$.  A general
theory of diffusion-controlled reactions with non-Markovian
binding/unbinding kinetics was developed in \cite{Grebenkov23b}.  The
binding events were described via a given distribution $\Psi(\ell)$ of
the threshold $\hat{\ell}$ for the boundary local time $\ell_t$.  When
$\Psi(\ell)$ is heavy tailed, i.e., $\Psi(\ell) \propto
(\ell_0/\ell)^{\alpha}$ at large $\ell$ with an exponent $0 <
\alpha < 1$, one deals with anomalous, non-Markovian binding kinetics.
Its competition with non-Markovian unbinding kinetics and their
consequences on diffusion-controlled reactions were described in
\cite{Grebenkov23b}.

\subsubsection*{Diffusion across permeable barriers}

Bressloff employed the encounter-based approach to describe diffusion
across permeable barriers
\cite{Bressloff22c,Bressloff23a,Bressloff23b}.  In this case, the
surface reaction is understood as permeation through the target set
$\Gamma$, and one can employ again the stopping condition $\ell_t <
\hat{\ell}$ as a trigger.  The subtle point here is that,
after permeation, the particle starts diffusing in the adjacent domain
(on the other side of $\Gamma$), whose encounter-based description
involves another boundary local time.  Indeed, after a number of
encounters with $\Gamma$ on the other side, the particle can permeate
back into the initial domain, and so on.  Note that Bressloff also
studied different functionals of reflected Brownian motion in the
presence of generalized surface reactions \cite{Bressloff22a}.

\subsubsection*{Extensions}

The encounter-based approach can be naturally extended in many ways.
As stated from the beginning, we focused on the simplest diffusive
dynamics in order to highlight the role of surface reactions.  The
main concepts of the encounter-based approach are expected to be valid
and valuable for much more general stochastic processes.  However, a
practical realization of such extensions requires considerable
mathematical efforts.  For instance, an extension to diffusion with a
gradient drift was discussed in \cite{Grebenkov22a}.  Further
extensions to anomalous diffusions such as continuous-time random
walks \cite{Metzler00,Klafter,Grebenkov10a,Grebenkov10b}, L\'evy
flights/walks \cite{Zaburdaev15}, diffusing-diffusivity processes
\cite{Chubynsky14,Sokolov17,Lanoiselee18}, and non-Markovian dynamics
\cite{Benichou14,Guerin16} present interesting perspectives.
Similarly, the encounter-based approach suggests an alternative way to
deal with intermittent diffusion with alternating phases of bulk and
surface diffusion
\cite{Levitz08,Chechkin09,Benichou10b,Benichou11b,Rojo11,Chechkin11,Chechkin12,Rupprecht12a,Rupprecht12b,Berezhkovskii15}
(see a review \cite{Benichou11} on intermittent search strategies).
Another extension consists in stochastic resetting of the position or
the boundary local time \cite{Bressloff22d,Benkhadaj22}.

\section{Conclusion} 
\label{sec:conclusion}

In this review, we presented the basic ideas and applications of the
encounter-based approach.  Unlike earlier publications on this topic,
we started from the discrete setting of a random walk on a lattice,
for which the formulation of the encounter-based approach is simple
and intuitively appealing.  In fact, the conventional propagator
$g_\sigma(\x,n|\x_0)$ that characterizes the random position of the
survived walker, is substituted by the joint distribution
$p(\x,L,n|\x_0)$ of the position and the number of encounters $L_n$ of
the particle with a given target set $\Gamma$.  This discrete setting
was then extended to Brownian motion that undergoes a first-order bulk
reaction inside the target set.  The number of encounters is naturally
replaced by the residence time $L_t$ in a bulk region $\Gamma$.  In
the next step, the target set is put on the boundary of an
impenetrable obstacle, in which case the residence time $L_t$ has to
be rescaled and replaced by the boundary local time $\ell_t$.  In all
three settings, the reaction event occurs when the (rescaled) number
of encounters ($L_n$, $L_t$ or $\ell_t$) exceeds a prescribed random
threshold.  The use of such a stopping condition generalizes
conventional reaction events with a constant reactivity, allowing one
to describe much more sophisticated processes such as activation or
passivation of the target set by interactions with the diffusing
particle.  The disentanglement of the diffusive dynamics from reaction
events is one of the crucial advantages of the encounter-based
approach.

Other advantages were rapidly illustrated by considering several
applications, including reversible binding/unbinding kinetics, the
escape problem, resource depletion by a population of diffusing
species, the statistics of various first-passage times for multiple
targets, etc.  We also showed how the Dirichlet-to-Neumann operator
can substitute the Laplace operator to compute the full propagator and
related quantities.  This spectral tool is particularly suitable for
describing diffusive explorations in the bulk between reflections on
the target set.

Despite impressive progress in this direction, many problems remain
open.  In particular, the numerical computation of the full propagator
in complex media is challenging that prohibits further understanding
of the impact of the geometrical complexity onto diffusion-controlled
reactions.  In the same vein, the disentanglement between the
diffusive dynamics and reactions event may help in solving
optimization problems such as finding an optimal structure of the
target set to minimize or maximize the moments of the reaction time,
or to reshape its distribution.  Such problems may appear in the
design of chemical reactors or in a programmable drug release in
pharmaceutical industry.  This research direction remains unexplored
yet.  Further developments of the encounter-based approach should
bring new insights onto the intricate diffusive dynamics in complex
environments and provide complementary views onto diffusion-controlled
reactions and other target problems.


\begin{thebibliography}{99.}%

\bibitem{Smoluchowski17}	M. Smoluchowski, 
				Versuch einer Mathematischen Theorie der Koagulations Kinetic Kolloider L\"osungen, 
				Z. Phys. Chem. {\bf 92U}, 129-168 (1918).


\bibitem{Rice}			S. A. Rice, 
				{\it Diffusion-limited reactions} 
				(Elsevier, Amsterdam, 1985).

\bibitem{Grebenkov23n}		D. S. Grebenkov,
				Diffusion-Controlled Reactions: An Overview
				(accepted to Molecules),



\bibitem{VanKampen}		N. G. Van Kampen,
				{\it Stochastic Processes in Physics and Chemistry}
				(Elsevier, Amsterdam, 1992).

\bibitem{Redner} 		S. Redner, 
				{\it A Guide to First Passage Processes}
				(Cambridge, Cambridge University press, 2001).

\bibitem{House}			J. E. House,
				{\it Principles of chemical kinetics}
				(Academic press, 2007).

\bibitem{Schuss}		Z. Schuss, 
				{\it Brownian Dynamics at Boundaries and Interfaces in Physics, Chemistry and Biology}
				(Springer, New York, 2013).

\bibitem{Metzler} 		R. Metzler, G. Oshanin, and S. Redner (Eds), 
				{\it First-Passage Phenomena and Their Applications}
				(Singapore, World Scientific, 2014).

\bibitem{Lindenberg}		K. Lindenberg, R. Metzler, and G. Oshanin (Eds),
				{\it Chemical Kinetics: Beyond the Textbook}
				(World Scientific, New Jersey, 2019).





\bibitem{Collins49}		F. C. Collins and G. E. Kimball, 
				Diffusion-controlled reaction rates, 
				J. Colloid Sci. {\bf 4}, 425-437 (1949).






\bibitem{Benichou00}		O. B\'enichou, M. Moreau, and G. Oshanin, 
				Kinetics of stochastically gated diffusion-limited reactions and geometry 
				of random walk trajectories, 
				Phys. Rev. E {\bf 61}, 3388-3406 (2000).

\bibitem{Reingruber09}		J. Reingruber and D. Holcman, 
				Gated Narrow Escape Time for Molecular Signaling, 
				Phys. Rev. Lett. {\bf 103}, 148102 (2009).

\bibitem{Lawley15}		S. D. Lawley and J. P. Keener,  
				A new derivation of Robin boundary conditions through homogenization of
				a stochastically switching boundary,
				SIAM J. Appl. Dyn. Syst. {\bf 14}, 1845-1867 (2015).



\bibitem{Berg77}		H. C. Berg and E. M. Purcell,
				Physics of chemoreception,
				Biophys. J. {\bf 20}, 193-219 (1977).

\bibitem{Berezhkovskii04}	A. M. Berezhkovskii, Y. A. Makhnovskii, M. I. Monine, V. Y. Zitserman, and S. Y. Shvartsman,
				Boundary homogenization for trapping by patchy surfaces,
				J. Chem. Phys. {\bf 121}, 11390 (2004).

\bibitem{Berezhkovskii06}	A. M. Berezhkovskii, M. I. Monine, C. B. Muratov, and S. Y. Shvartsman,
				Homogenization of boundary conditions for surfaces with regular arrays of traps,
				J. Chem. Phys. {\bf 124}, 036103 (2006).

\bibitem{Muratov08}		C. Muratov and S. Shvartsman, 
				Boundary homogenization for periodic arrays of absorbers,
				Multiscale Model. Simul. {\bf 7}, 44-61 (2008).

\bibitem{Bernoff18}		A. Bernoff, A. Lindsay, and D. Schmidt, 
				Boundary homogenization and capture time distributions of semipermeable membranes 
				with periodic patterns of reactive sites,
				Multiscale Model. Simul. {\bf 16}, 1411-1447 (2018).

\bibitem{Grebenkov19b}		D. S. Grebenkov, 
				Spectral theory of imperfect diffusion-controlled reactions on heterogeneous catalytic surfaces, 
				J. Chem. Phys. {\bf 151}, 104108 (2019).

\bibitem{Punia21}		B. Punia, S. Chaudhury, and A. B. Kolomeisky,
				Understanding the Reaction Dynamics on Heterogeneous Catalysts Using a Simple Stochastic Approach,
				J. Phys. Chem. Lett. {\bf 12}, 11802-11810 (2021).



\bibitem{Weiss86}		G. H. Weiss, 
				Overview of theoretical models for reaction rates,
				J. Stat. Phys. {\bf 42}, 3-36 (1986).

\bibitem{Hanggi90}		P. H\"anggi, P. Talkner, and M. Borkovec, 
				Reaction-rate theory: fifty years after Kramers, 
				Rev. Mod. Phys. {\bf 62}, 251-341 (1990).

\bibitem{Zhou91}		H.-X. Zhou and R. Zwanzig, 
				A rate process with an entropy barrier,
				J. Chem. Phys. {\bf 94}, 6147-6152 (1991).

\bibitem{Reguera06}		D. Reguera, G. Schmid, P. S. Burada, J.-M. Rub{\'\i}, P. Reimann, and P. H\"anggi, 
				Entropic Transport: Kinetics, Scaling, and Control Mechanisms,
				Phys. Rev. Lett. {\bf 96}, 130603 (2006).

\bibitem{Chapman16}		S. J. Chapman, R. Erban, and S. Isaacson,
				Reactive boundary conditions as limits of interaction potentials 
				for Brownian and Langevin dynamics,
				SIAM J. Appl. Math {\bf 76}, 368-390 (2016).




\bibitem{Grebenkov06}		D. S. Grebenkov, 
				Partially Reflected Brownian Motion: A Stochastic Approach to Transport Phenomena, 
				in ``Focus on Probability Theory'', Ed. L. R. Velle, pp. 135-169 
				(Nova Science Publishers, New York, 2006). 

\bibitem{Grebenkov07}		D. S. Grebenkov, 
				Residence times and other functionals of reflected Brownian motion, 
				Phys. Rev. E {\bf 76}, 041139 (2007).

\bibitem{Erban07}		R. Erban and S. J. Chapman,
				Reactive boundary conditions for stochastic simulations of reaction-diffusion processes,
				Phys. Biol. {\bf 4}, 16-28 (2007).

\bibitem{Singer08}		A. Singer, Z. Schuss, A. Osipov, and D. Holcman, 
				Partially reflected diffusion,
				SIAM J. Appl. Math. {\bf 68}, 844-868 (2008).

\bibitem{Grebenkov20}		D. S. Grebenkov, 
				Paradigm Shift in Diffusion-Mediated Surface Phenomena, 
				Phys. Rev. Lett. {\bf 125}, 078102 (2020).

\bibitem{Piazza22}		F. Piazza,
				The physics of boundary conditions in reaction-diffusion problems,
				J. Chem. Phys. {\bf 157}, 234110 (2022).

\bibitem{Grebenkov23b}		D. S. Grebenkov, 
				Diffusion-controlled reactions with non-Markovian binding/unbinding kinetics, 
				J. Chem. Phys. {\bf 158}, 214111 (2023). 





\bibitem{Sano79}		H. Sano and M. Tachiya, 
				Partially diffusion-controlled recombination, 
				J. Chem. Phys. {\bf 71}, 1276-1282 (1979).

\bibitem{Brownstein79}		K. R. Brownstein and C. E. Tarr, 
				Importance of Classical Diffusion in NMR Studies of Water in Biological Cells, 
				Phys. Rev. A {\bf 19}, 2446-2453 (1979).


\bibitem{Powles92}		J. G. Powles, M. J. D. Mallett, G. Rickayzen, and W. A. B. Evans, 
				Exact analytic solutions for diffusion impeded by an infinite array of partially 
				permeable barriers, 
				Proc. R. Soc. London A {\bf 436}, 391-403 (1992).

\bibitem{Sapoval94}		B. Sapoval, 
				General Formulation of Laplacian Transfer Across Irregular Surfaces, 
				Phys. Rev. Lett. {\bf 73}, 3314-3316 (1994).

\bibitem{Sapoval02}		B. Sapoval, M. Filoche, and E. Weibel, 
				Smaller is better -- but not too small: A physical scale for the design 
				of the mammalian pulmonary acinus, 
				Proc. Nat. Ac. Sci. USA {\bf 99}, 10411-10416 (2002).

\bibitem{Grebenkov05}		D. S. Grebenkov, M. Filoche, B. Sapoval, and M. Felici, 
				Diffusion-reaction in Branched Structures: Theory and Application to the Lung Acinus, 
				Phys. Rev. Lett. {\bf 94}, 050602 (2005).

\bibitem{Traytak07}		S. D. Traytak and W. Price, 
				Exact solution for anisotropic diffusion-controlled reactions with partially reflecting conditions, 
				J. Chem. Phys. {\bf 127}, 184508 (2007).

\bibitem{Bressloff08}		P. C. Bressloff, B. A. Earnshaw, and M. J. Ward, 
				Diffusion of protein receptors on a cylindrical dendritic membrane with partially absorbing traps, 
				SIAM J. Appl. Math. {\bf 68}, 1223-1246 (2008).


\bibitem{Grebenkov15}		D. S. Grebenkov, 
				Analytical representations of the spread harmonic measure density,
				Phys. Rev. E {\bf 91}, 052108 (2015).

\bibitem{Serov16}		A. S. Serov, C. Salafia, D. S. Grebenkov, and M. Filoche, 
				The Role of Morphology in Mathematical Models of Placental Gas Exchange, 
				J. Appl. Physiol. {\bf 120}, 17-28 (2016).


\bibitem{Grebenkov17a}		D. S. Grebenkov and G. Oshanin, 				
				Diffusive escape through a narrow opening: new insights into a classic problem,
				Phys. Chem. Chem. Phys. {\bf 19}, 2723-2739 (2017).


\bibitem{Grebenkov18}		D. S. Grebenkov, R. Metzler, and G. Oshanin, 
				Strong defocusing of molecular reaction times results from an 
				interplay of geometry and reaction control, 
				Commun. Chem. {\bf 1}, 96 (2018).

\bibitem{Grebenkov18b}		D. S. Grebenkov, R. Metzler, and G. Oshanin,
				Towards a full quantitative description of single-molecule reaction kinetics in biological cells,
				Phys. Chem. Chem. Phys. {\bf 20}, 16393-16401 (2018).  

\bibitem{Grebenkov19c}		D. S. Grebenkov, R. Metzler, and G. Oshanin, 
				Full distribution of first exit times in the narrow escape problem, 
				New J. Phys. {\bf 21}, 122001 (2019).

\bibitem{Guerin21}		T. Gu\'erin, M. Dolgushev, O. B\'enichou, and R. Voituriez,
				Universal kinetics of imperfect reactions in confinement,
				Commun. Chem. {\bf 4}, 157 (2021).

\bibitem{Piazza19}		F. Piazza and D. S. Grebenkov, 
				Diffusion-controlled reaction rate on non-spherical partially absorbing axisymmetric surfaces, 
				Phys. Chem. Chem. Phys. {\bf 21}, 25896-25906 (2019).



\bibitem{Gardiner}		C. W. Gardiner,
				{\it Handbook of stochastic methods for physics, chemistry and the natural sciences}
				(Springer: Berlin, 1985).

\bibitem{Grebenkov13}		D. S. Grebenkov and B.-T. Nguyen, 
				Geometrical structure of Laplacian eigenfunctions, 
				SIAM Rev. {\bf 55}, 601-667 (2013).









\bibitem{Grebenkov19a}		D. S. Grebenkov, 
				Probability distribution of the boundary local time of reflected Brownian 
				motion in Euclidean domains, 
				Phys. Rev. E {\bf 100}, 062110 (2019).

\bibitem{Grebenkov20b}		D. S. Grebenkov, 
				Surface Hopping Propagator: An Alternative Approach to Diffusion-Influenced Reactions, 
				Phys. Rev. E {\bf 102}, 032125 (2020).

\bibitem{Grebenkov20c}		D. S. Grebenkov, 
				Joint distribution of multiple boundary local times and related first-passage 
				time problems with multiple targets, 
				J. Stat. Mech. 103205 (2020).

\bibitem{Grebenkov21}		D. S. Grebenkov, 
				Statistics of boundary encounters by a particle diffusing outside a compact planar domain, 
				J. Phys. A.: Math. Theor. {\bf 54}, 015003 (2021). 

\bibitem{Grebenkov22a}		D. S. Grebenkov, 
				An encounter-based approach for restricted diffusion with a gradient drift, 
				J. Phys. A: Math. Theor. {\bf 55}, 045203 (2022).

\bibitem{Bressloff22d}		P. C. Bressloff,
				Diffusion-mediated surface reactions and stochastic resetting,
				J. Phys. A: Math. Theor. {\bf 55}, 275002 (2022).

\bibitem{Benkhadaj22}		Z. Benkhadaj and D. S. Grebenkov, 
				Encounter-based approach to diffusion with resetting, 
				Phys. Rev. E {\bf 106}, 044121 (2022).

\bibitem{Grebenkov22b}		D. S. Grebenkov, 
				Depletion of Resources by a Population of Diffusing Species, 
				Phys. Rev. E {\bf 105}, 054402 (2022). 

\bibitem{Bressloff22b}		P. C. Bressloff,
				Narrow capture problem: an encounter-based approach to partially reactive targets,
				Phys. Rev. E {\bf 105}, 034141 (2022).

\bibitem{Grebenkov22c}		D. S. Grebenkov, 
				Statistics of diffusive encounters with a small target: Three complementary approaches, 
				J. Stat. Mech. 083205 (2022). 

\bibitem{Bressloff22a}		P. C. Bressloff,
				Diffusion-mediated absorption by partially-reactive targets: 
				Brownian functionals and generalized propagators,
				J. Phys. A: Math. Theor. {\bf 55}, 205001 (2022).

\bibitem{Grebenkov23a}		D. S. Grebenkov, 
				Encounter-based approach to the escape problem, 
				Phys. Rev. E {\bf 107}, 044105 (2023).

\bibitem{Bressloff22c}		P. C. Bressloff, 
				A probabilistic model of diffusion through a semipermeable barrier,
				Proc. Roy. Soc. A {\bf 478}, 20220615 (2022).

\bibitem{Bressloff23a}		P. C. Bressloff,
				Renewal equation for single-particle diffusion through a semipermeable interface,
				Phys. Rev. E. {\bf 107}, 014110 (2023).

\bibitem{Bressloff23b}		P. C. Bressloff,
				Renewal equations for single-particle diffusion in multilayered media,
				SIAM J. Appl. Math. {\bf 83}, 1518-1545 (2023).



\bibitem{Rubin82}		R. J. Rubin and G. H. Weiss, 
				Random walks on lattices. The problem of visits to a set of points revisited,
				J. Math. Phys. {\bf 23}, 250-253 (1982).


\bibitem{Feller}		W. Feller, 
				{\it An Introduction to Probability Theory and Its Applications} 
				(John Wiley, New York, 1968).

\bibitem{Spitzer}		F. Spitzer,
				{\it Principles of Random Walk} 
				(New York: Springer, 1976).

\bibitem{Hughes}		B. D. Hughes,  
				{\it Random Walks and Random Environments}
				(Clarendon Press, Oxford, 1995). 



\bibitem{Montroll65}		E. W. Montroll and G. H. Weiss,
				Random Walks on Lattices. II,
				J. Math. Phys. {\bf 6}, 167-181 (1965).


\bibitem{Guttmann10}		A. J. Guttmann,
				Lattice Green's functions in all dimensions,
				J. Phys. A: Math. Theor. {\bf 43}, 305205 (2010).

\bibitem{Joyce98}		G. S. Joyce,
				Exact evaluation of the simple cubic lattice Green function for a general lattice point,
				J. Phys. A: Math. Gen. {\bf 31}, 5105-5115 (1998).



\bibitem{Szabo84}		A. Szabo, G. Lamm, and G. H. Weiss,
				Localized partial traps in diffusion processes and random walks,
				J. Stat. Phys. {\bf 34}, 225-238 (1984).




\bibitem{Borodin}		A. N. Borodin and P. Salminen,
				{\it Handbook of Brownian Motion: Facts and Formulae}
				(Birkhauser Verlag, Basel-Boston-Berlin, 1996).

\bibitem{Majumdar05}		S. N. Majumdar,
				Brownian functionals in physics and computer science,
				Curr. Sci. {\bf 88}, 2076-2092 (2005).


\bibitem{Ito}			K. It\^o and H. P. McKean, 
				{\it Diffusion Processes and Their Sample Paths}
				(Berlin: Springer, 1965).

\bibitem{Freidlin}		M. Freidlin,
				{\it Functional Integration and Partial Differential Equations}
				(Annals of Mathematics Studies, Princeton, NJ: Princeton University Press, 1985).



\bibitem{Levitin}		M. Levitin, D. Mangoubi, and I. Polterovich,
				Topics in Spectral Geometry
				(Preliminary version, May 29, 2023; https://www.michaellevitin.net/Book/TSG230529.pdf)


\bibitem{Grebenkov06b}		D. S. Grebenkov, 
				Scaling Properties of the Spread Harmonic Measures,
				Fractals {\bf 14}, 231-243 (2006).










\bibitem{Weiss83}		G. H. Weiss, K. E. Shuler, and K. Lindenberg,
				Order Statistics for First Passage Times in Diffusion Processes,
				J. Stat. Phys. {\bf 31}, 255-278 (1983).

\bibitem{Bray13}		A. J. Bray, S. N. Majumdar and G. Schehr,
				Persistence and first-passage properties in non-equilibrium systems,
				Adv. Phys. {\bf 62}, 225-361 (2013).

\bibitem{Schuss19}		Z. Schuss, K. Basnayake and D. Holcman,
				Redundancy principle and the role of extreme statistics in molecular and cellular biology,
				Phys. Life Rev. {\bf 28}, 52-79 (2019).

\bibitem{Basnayake19}		K. Basnayake, Z. Schuss and D. Holcman,
				Asymptotic formulas for extreme statistics of escape times in 1, 2 and 3-dimensions,
				J. Nonlinear Sci. {\bf 29} 461-499 (2019).
				
\bibitem{Lawley20}		S. D. Lawley,
				Distribution of extreme first passage times of diffusion,
				J. Math. Biol. {\bf 80}, 2301-2325 (2020).				

\bibitem{Lawley20b}		S. D. Lawley and J. B. Madrid,
				A probabilistic approach to extreme statistics of Brownian escape times in dimensions 1, 2, and 3,
				J. Nonlinear Sci. {\bf 30}, 1207-1227 (2020).

\bibitem{Lawley20c}		S. D. Lawley,
				Universal formula for extreme first passage statistics of diffusion,
				Phys. Rev. E {\bf 101}, 012413 (2020).

\bibitem{Madrid20}		J. B. Madrid and S. D. Lawley,
				Competition between slow and fast regimes for extreme first passage times of diffusion,
				J. Phys. A: Math. Theor. {\bf 53}, 335002 (2020).

\bibitem{Grebenkov20a}		D. S. Grebenkov, R. Metzler, and G. Oshanin,
				From single-particle stochastic kinetics to macroscopic reaction rates: 
				fastest first-passage time of $N$ random walkers,
				New J. Phys. {\bf 22}, 103004 (2020).

\bibitem{Grebenkov22}		D. S. Grebenkov, R. Metzler, and G. Oshanin,
				Search efficiency in the Adam-Delbr{\"u}ck reduction-of-dimensionality scenario versus direct diffusive search,
				New J. Phys. {\bf 24}, 083035 (2022).


\bibitem{Lauffenburger}		D. A. Lauffenburger and J. Linderman, 
				{\it Receptors: Models for Binding, Trafficking, and Signaling}
				(Oxford University Press, Oxford, 1993).



\bibitem{Schuss07}		Z. Schuss, A. Singer, and D. Holcman,
				The narrow escape problem for diffusion in cellular microdomains,
				Proc. Nat. Acad. Sci. USA {\bf 104}, 16098-16103 (2007).

\bibitem{Benichou08}		O. B\'enichou and R. Voituriez,
				Narrow-escape time problem: time needed for a particle to exit a confining 
				domain through a small window,
				Phys. Rev. Lett. {\bf 100}, 168105 (2008).

\bibitem{Holcman14}		D. Holcman and Z. Schuss, 
				The narrow escape problem,
				SIAM Rev. {\bf 56}, 213-257 (2014).





\bibitem{Metzler00}		R. Metzler and J. Klafter,
				The random walk's guide to anomalous diffusion: a fractional dynamics approach,
				Phys. Rep. {\bf 339}, 1-77 (2000).

\bibitem{Klafter}		J. Klafter and I. M. Sokolov,
				First Steps in Random Walks: From Tools to Applications
				(Oxford University Press, 2011).

\bibitem{Grebenkov10a}		D. S. Grebenkov, 
				Searching for partially reactive sites: Analytical results for spherical targets, 
				J. Chem. Phys. {\bf 132}, 034104 (2010).

\bibitem{Grebenkov10b}		D. S. Grebenkov, 
				Subdiffusion in a bounded domain with a partially absorbing-reflecting boundary, 
				Phys. Rev. E {\bf 81}, 021128 (2010).

\bibitem{Zaburdaev15}		V. Zaburdaev, S. Denisov, and J. Klafter,
				L\'evy walks,
				Rev. Mod. Phys. {\bf 87}, 483-530 (2015).


\bibitem{Chubynsky14}		M. V. Chubynsky and G. W. Slater,
				Diffusing Diffusivity: A Model for Anomalous, yet Brownian, Diffusion,
				Phys. Rev. Lett. {\bf 113}, 098302 (2014). 

\bibitem{Sokolov17}		A. V. Chechkin, F. Seno, R. Metzler, and I. M. Sokolov,
				Brownian yet Non-Gaussian Diffusion: From Superstatistics to Subordination of Diffusing Diffusivities,
				Phys. Rev. X {\bf 7}, 021002 (2017).

\bibitem{Lanoiselee18}		Y. Lanoisel\'ee, N. Moutal, and D. S. Grebenkov, 
				Diffusion-limited reactions in dynamic heterogeneous media, 
				Nature Commun. {\bf 9}, 4398 (2018).


\bibitem{Benichou14}		O. B\'enichou and R. Voituriez,
				From first-passage times of random walks in confinement to geometry-controlled kinetics,
				Phys. Rep. {\bf 539}, 225-284 (2014).

\bibitem{Guerin16}		T. Gu\'erin, N. Levernier, O. B\'enichou, and R. Voituriez,
				Mean first-passage times of non-Markovian random walkers in confinement,
				Nature {\bf 534}, 356-359 (2016).






\bibitem{Levitz08}		P. Levitz, M. Zinsmeister, P. Davidson, D. Constantin, and O. Poncelet,
				Intermittent Brownian dynamics over a rigid strand: 
				Heavily tailed relocation statistics in a simple geometry,
				Phys. Rev. E {\bf 78}, 030102(R) (2008).

\bibitem{Chechkin09}		A. V. Chechkin, I. M. Zaid, M. Lomholt, I. M. Sokolov and R. Metzler, 
				Bulk-mediated surface diffusion along a cylinder: Propagators and crossovers, 
				Phys. Rev. E {\bf 79}, 040105(R) (2009).

\bibitem{Benichou10b}		O. B\'enichou, D. S. Grebenkov, P. Levitz, C. Loverdo and R. Voituriez, 
				Optimal Reaction Time for Surface-Mediated Diffusion,
				Phys. Rev. Lett. {\bf 105}, 150606 (2010).


\bibitem{Benichou11b}		O. B\'enichou, D. S. Grebenkov, P. Levitz, C. Loverdo and R. Voituriez, 
				Mean first-passage time of surface-mediated diffusion in spherical domains,
				J. Stat. Phys. {\bf 142}, 657-685 (2011).

\bibitem{Rojo11}		F. Rojo and C. E. Budde, 
				Enhanced diffusion through surface excursion: A master-equation approach 
				to the narrow-escape-time problem,
				Phys. Rev. E {\bf 84}, 021117 (2011).

\bibitem{Chechkin11}		A. V. Chechkin, I. M. Zaid, M. Lomholt, I. M. Sokolov, and R. Metzler, 
	 			Effective surface motion on a reactive cylinder of particles that perform intermittent 
				bulk diffusion, 
				J. Chem. Phys. {\bf 134}, 204116 (2011).

\bibitem{Chechkin12}		A. V. Chechkin, I. M. Zaid, M. Lomholt, I. M. Sokolov, and R. Metzler, 
				Bulk-mediated diffusion on a planar surface: Full solution, 
				Phys. Rev. E {\bf 86}, 041101 (2012).

\bibitem{Rupprecht12a}		J.-F. Rupprecht, O. B\'enichou, D. S. Grebenkov and R. Voituriez,
				Exact mean exit time for surface-mediated diffusion,
				Phys. Rev. E {\bf 86}, 041135 (2012).

\bibitem{Rupprecht12b}		J.-F. Rupprecht, O. B\'enichou, D. S. Grebenkov and R. Voituriez,
				Kinetics of active surface-mediated diffusion in spherically symmetric domains,
				J. Stat. Phys. {\bf 147}, 891-918 (2012).

\bibitem{Berezhkovskii15}	A. M. Berezhkovskii, L. Dagdug, and S. M. Bezrukov, 
				A new approach to the problem of bulk-mediated surface diffusion, 
				J. Chem. Phys. {\bf 143}, 084103 (2015).


\bibitem{Benichou11}		O. B\'enichou, C. Loverdo, M. Moreau, and R. Voituriez,
				Intermittent search strategies,
				Rev. Mod. Phys. {\bf 83}, 81-130 (2011).






\end{thebibliography}
\end{document}